 \documentclass[final,5p,times,twocolumn]{elsarticle}

\usepackage{graphicx}
\usepackage{wrapfig}

\usepackage{lipsum}
\usepackage{dirtytalk}
\usepackage[utf8]{inputenc}
\usepackage{subfigure}
\usepackage{mathtools}
\usepackage[algo2e,linesnumbered]{algorithm2e} 
\usepackage{algorithmic}
\usepackage{algorithm}
\usepackage{dirtytalk}
\usepackage[utf8]{inputenc}
\usepackage[english]{babel}
\usepackage{amsthm}
\usepackage{amsmath}
\usepackage{lipsum}
\usepackage{titlesec}

\DeclarePairedDelimiter\ceil{\lceil}{\rceil}
\DeclarePairedDelimiter\floor{\lfloor}{\rfloor}

\usepackage[colorlinks]{hyperref}
\newtheorem{theorem}{Theorem}
\newtheorem{lemma}{Lemma}
\theoremstyle{definition}
\newtheorem{definition}{Definition}


\showoutput
\titlespacing\section{0pt}{4pt plus 2pt minus 2pt}{4pt plus 2pt minus 2pt}
\titlespacing\subsection{0pt}{3pt plus 2pt minus 2pt}{3pt plus 2pt minus 2pt}
\titlespacing\subsubsection{0pt}{2pt plus 2pt minus 2pt}{2pt plus 2pt minus 2pt}
\usepackage{bbding}
\journal{Ad Hoc Networks ELSEVIER}
\begin{document}
%\long\def\/*#1*/{}
%\setlength{\abovedisplayskip}{0.25pt}
%\setlength{\belowdisplayskip}{0.25pt}
%% For including figures, graphicx.sty has been loaded in
%% elsarticle.cls. If you prefer to use the old commands
%% please give \usepackage{epsfig}

%% The amssymb package provides various useful mathematical symbols
%\usepackage{amssymb}
%% The amsthm package provides extended theorem environments
%% \usepackage{amsthm}

%% The lineno packages adds line numbers. Start line numbering with
%% \begin{linenumbers}, end it with \end{linenumbers}. Or switch it on
%% for the whole article with \linenumbers.
%% \usepackage{lineno}
\begin{frontmatter}



\title{Efficient Medium Access Arbitration Among Interfering WBANs Using Latin Rectangles}

\author[label1]{Mohamad Jaafar Ali}
\address[label1]{LIPADE, University of Paris Descartes, Sorbonne Paris Cit\'{e}, Paris, Frane}
\ead{mohamadjali84@gmail.com}
\author[label1,label2]{Hassine Moungla}
\address[label2]{UMR 5157, CNRS, Institute Mines-Telecom, T\'{e}l\'{e}com SudParis, Nano-Innov CEA Saclay, France}
\ead{hassine.moungla@parisdescartes.fr}
\author[label3]{Mohamed Younis}
\address[label3]{Department of Computer Science and Electrical Engineering, University of Maryland, Baltimore County, United States}
\ead{younis@umbc.edu}
\author[label1]{Ahmed Mehaoua}
\ead{ahmed.mehaoua@parisdescartes.fr}

 %\author[label1,label2]{}

\begin{abstract}
The overlap of transmission ranges among multiple Wireless Body Area Networks (\textit{WBAN}s) is referred to as coexistence. The interference is most likely to affect the communication links and degrade the performance when sensors of different \textit{WBAN}s simultaneously transmit using the same channel. In this paper, we propose a distributed approach that adapts to the size of the network, i.e., the number of coexisting \textit{WBAN}s, and to the density of sensors forming each individual \textit{WBAN} in order to minimize the impact of co-channel interference through dynamic channel hopping based on Latin rectangles. Furthermore, the proposed approach opts to reduce the overhead resulting from channel hopping, and lowers the transmission delay, and saves the power resource at both sensor- and \textit{WBAN}-levels. Specifically, we propose two schemes for channel allocation and medium access scheduling to diminish the probability of inter-\textit{WBAN} interference. The first scheme, namely, Distributed Interference Avoidance using Latin rectangles (\textit{DAIL}), assigns channel and time-slot combination that reduces the probability of medium access collision. \textit{DAIL} suits crowded areas, e.g., high density of coexisting \textit{WBAN}s, and involves overhead due to frequent channel hopping at the \textit{WBAN} coordinator and sensors. The second scheme, namely, \textit{CHIM}, takes advantage of the relatively lower density of collocated \textit{WBAN}s to save power by hopping among channels only when interference is detected at the level of the individual nodes. We present an analytical model that derives the collision probability and network throughput. The performance of \textit{DAIL} and \textit{CHIM} is further validated through simulations.
\end{abstract}
\begin{keyword}
WBAN \sep Interference \sep Channel Hopping \sep TDMA \sep Latin Square
\end{keyword}
\end{frontmatter}

%\section{}
%\label{}
\section{Introduction}
\subsection{WBANs Overview}
The recent technological advances in wireless communication, microelectronics have enabled the development of low-power, intelligent, miniaturized sensor nodes to be implanted in or attached to the human bodies. Inter-networking these exciting and evolving devices is referred to as a \textit{WBAN} and is revolutionizing remote health monitoring and telemedicine in general. In essence, a \textit{WBAN} is a wireless short-range communication network consisting of a single coordinator and a finite number of low-power wireless sensor devices. These sensors enable continual monitoring of the physiological state of the body in stationary or mobility scenarios. A \textit{WBAN} also includes a coordinator that collects the measurements of the individual sensors and shares them with a remote monitoring station (command canter) \cite{key1}. 

\textit{WBAN}s can serve in various applications such as ubiquitous health care, telemedicine, entertainment, sports and military \cite{key3}. Basically, the \textit{WBAN} sensors may observe the heart (electrocardiography) and the brain electrical (electroencephalographs) activities as well as vital signs and parameters like blood sugar, blood pressure, body temperature, oxygen saturation, \textit{CO2} concentration, etc. The collected data can then be forwarded to a hospital for medical assessment.

\begin{figure}
  \centering
   \includegraphics[width=0.275\textwidth,height=0.2\textheight]{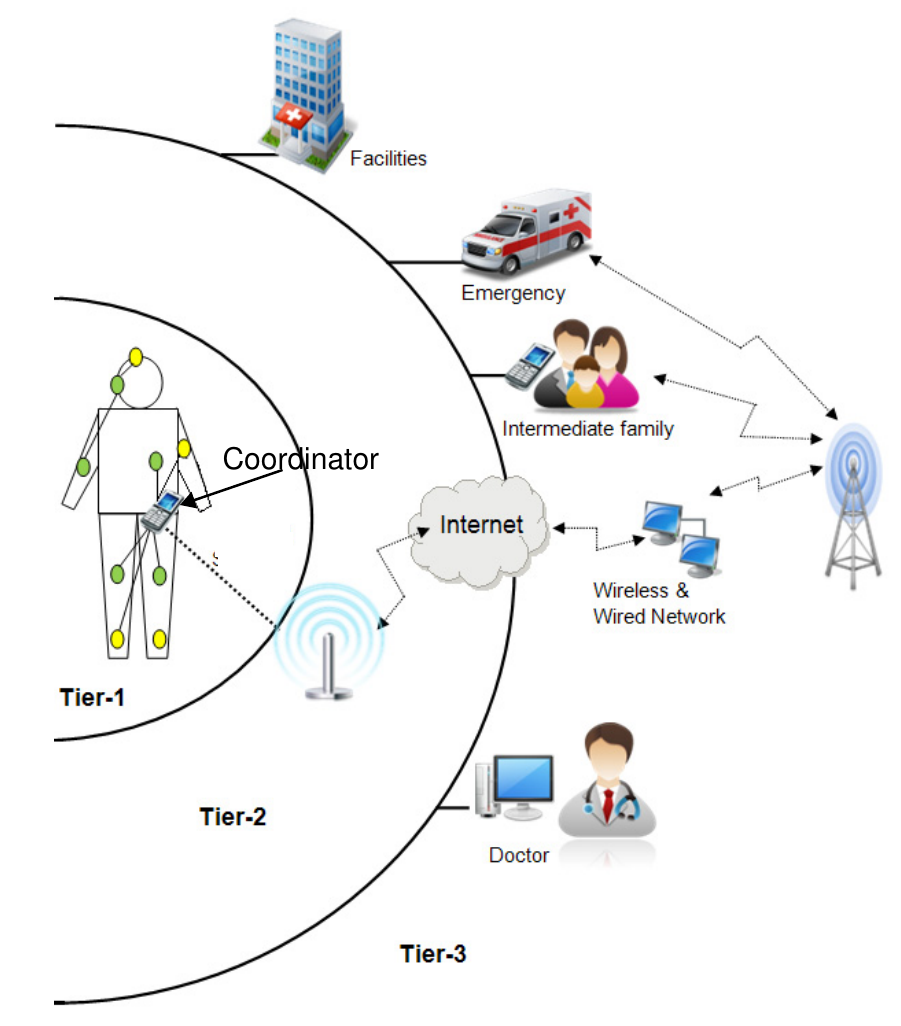}
\caption{Communication architecture of \textit{WBAN}s}
\label{trs}
\end{figure}

\textit{WBAN}s are becoming increasingly pervasive; their coexistence will become a serious issue in the upcoming years. Eleven million sensors were estimated to be used in 2009 and are predicted to reach 485 million by 2018 \cite{key1}. The communication architecture of involving \textit{WBAN}s composes of three tiers as illustrated in \textbf{Figure \ref{trs}}.
\begin{itemize}
\item \textbf{\textit{Tier-1:}} Intra-\textit{WBAN} communication – This tier reflects the interaction of the \textit{WBAN} nodes (sensors and coordinator). Sensors forward their measurements to the coordinator. The coordinator processes the collected physiological data and then transmits to an access point in Tier-2
\item \textbf{\textit{Tier-2:}} Inter-\textit{WBAN} communication – This communication tier is between the coordinator and one or more access points. Tier-2 communication aims to interconnect \textit{WBAN}s with various networks (cellular, etc,)
\item \textbf{\textit{Tier-3:}} Caregiver – This tier reflects medical facility, nurses, doctors, and family members, and can be viewed as remote command centers
\end{itemize}
\subsection{Co-Channel Interference amongst WBANs}
Due to their social nature, the individual \textit{WBAN}s may move towards each other in crowded areas such as a hospital's lobby. The interference may occur because of the smaller number of channels, i.e., $16$ in \textit{ZIGBEE} \cite{key5a},\cite{key5} than the number of coexisting \textit{WBAN}s. Even when few \textit{WBAN}s coexist, such interference may affect the communication links by decreasing the signal-to-interference-plus-noise-ratio (\textit{SINR}) of the received signal and thus may have an adverse impact on the reliability of \textit{WBAN}s. The overlap among the transmission ranges of multiple collocated \textit{WBAN}-based networks may cause medium access collision. Basically, a link in one \textit{WBAN} will suffer interference due to transmissions on the same channel made by a device mounted on another person in the vicinity. Thus, the inter-\textit{WBAN} interference may affect the communication links and degrade the performance as well as increase the packet loss of each individual \textit{WBAN}. In addition, inter-\textit{WBAN} interference drains the most valuable resource of \textit{WBAN}'s sensor, namely, power, while trying to compete for a better signal-to-noise-ratio (SNR); it also leads to throughput degradation. Thus, the co-channel interference becomes one of the most important problems that happen when multiple independent and uncoordinated \textit{WBAN}s work concurrently in the close proximity of each other. Therefore, interference mitigation is of very importance to improve the performance of the whole network \cite{key3,key2}.
\begin{figure}
  \centering
   \includegraphics[width=0.25\textwidth,height=0.15\textheight]{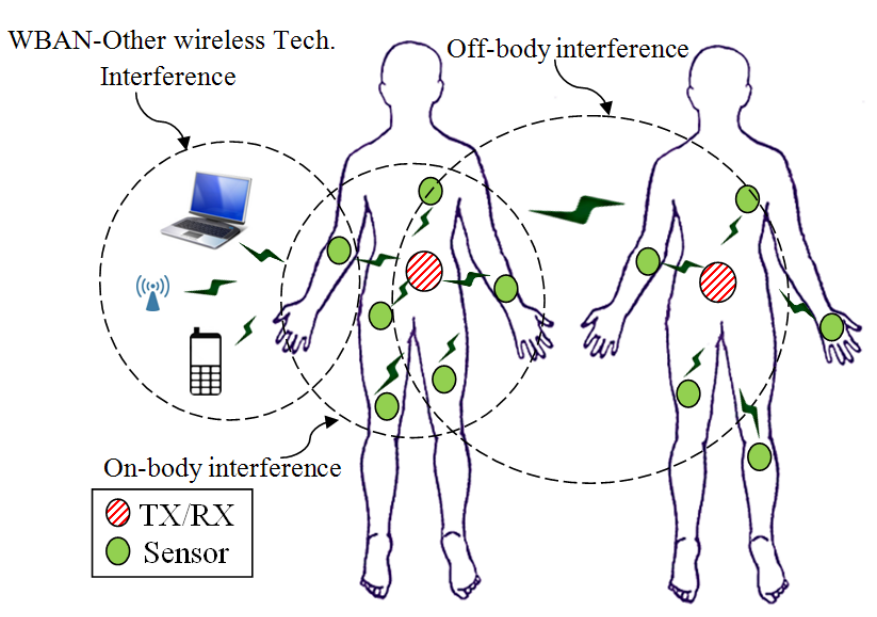}
\caption{Radio co-channel interference between \textit{WBAN}s and other wireless networks}
\label{fig_1_2}
\end{figure}
In addition, the resource constrained nature of \textit{WBAN}s in terms of limited power supply, sensor's and antenna's size, the transmission range, and the quality of service (QoS) requirements make the application of advanced antenna and power control mechanisms that proved their efficiency in interference mitigation in cellular and wireless sensor networks unsuitable for \textit{WBAN}s. The protocols proposed for these networks do not consider the stringent properties of \textit{WBAN}s, for instance, power control mechanisms are not suitable for \textit{WBAN}s because they require different levels of transmission power, whilst, longer lifetime of sensors batteries is required. The simple design and shape of sensor's antenna make signal processing very hard in \textit{WBAN}s because of the inhomogeneous nature of the human body characterized by high signal attenuation and distortion. Due to the mobile nature of \textit{WBAN}s, the \textit{WBAN}s may change their position relative to each other and the individual sensors in the same \textit{WBAN} may change their location relative to each other. Such dynamic nature and the absence of coordination as well as the synchronization amongst \textit{WBAN}s, make the allocation of a global centralized entity to manage \textit{WBAN}s coexistence and mitigate the interference unsuitable for \textit{WBAN}s \cite{key3, key4}. 

Recently, the \textit{IEEE 802.15.6} standard \cite{key5} has emerged as the defacto medium access control protocol for \textit{WBAN}s. The standard requires the system to function properly within the transmission range of up to 3 meters when up to 10 \textit{WBAN}s are collocated. It also has to support 60 sensors in a $6m^{3}$ space (256 sensors in a $3m^{3}$). The standard recommends the use of time division multiple access schemes (TDMA) as an alternative solution to avoid co-channel interference within \textit{WBAN}s. To this end, it also proposes three mechanisms for inter-\textit{WBAN} co-channel interference mitigation, namely, beacon shifting, active superframe interleaving and channel hopping \cite{key4, key6}, where a superframe is defined as a time period consisting of two successive frames, an active frame dedicated for sensors transmissions and an inactive period dedicated for coordinators transmissions. 
\begin{itemize}
\item \textbf{Beacon shifting :} a \textit{WBAN}'s coordinator may transmit its beacons at different time offsets by including a shifting sequence field in its beacons that is not used by neighbor coordinators.
\item \textbf{Superframe interleaving :} a \textit{WBAN}'s coordinator exchanges information with its counterparts in other \textit{WBAN}s in order to make the active period of its superframe not to overlap with the active periods of other superframes.
\item \textbf{Channel hopping :} a coordinator may change its operating channel periodically by choosing a particular channel hopping sequence that should not be used by nearby \textit{WBAN}s.
\end{itemize}
However, the first two mechanisms require distributed guard time computation that makes time synchronization very hard to achieve across \textit{WBAN}s. Therefore, we exploit the unlicensed international spectrum available in the \textit{ZIGBEE} standard and pursue the approach of spectrum allocation to resolve the problem of inter-\textit{WBAN} interference mitigation not only through the channel but also the channel to time-slot hopping.
\subsection{Contribution and Organization}
Due to their social and highly mobile nature, \textit{WBAN}s may coexist in the proximity of each other and due to the absence of coordination amongst them, the active periods of the corresponding superframes may overlap with each other, hence the co-channel interference may happen. Particularly, when some sensors of the interfering \textit{WBAN}s simultaneously access the same channel, the corresponding transmissions of these sensors experience medium access collision. In this paper, we address this problem using Latin squares and propose two schemes for channel allocation and medium access scheduling to diminish the probability of inter-\textit{WBAN} interference. The first scheme, namely, \textit{DAIL}, assigns channel and time-slot combinations that reduce the probability of medium access collision. No inter-\textit{WBAN} coordination is needed in \textit{DAIL}. Despite being very effective, \textit{DAIL} involves overhead due to frequent channel hopping at the coordinator and the sensors. The second scheme, namely, \textit{CHIM}, takes advantage of the relatively lower density of collocated \textit{WBAN}s and pursues hopping among channels only when interference is detected in order to save power. In summary, we contribute the following:
\begin{itemize}
\item \textit{DAIL, a distributed scheme which enables time-based channel hopping using Latin rectangles in order to minimize the medium access collision and avoid the co-channel interference amongst coexisting \text{WBAN}s.}
\item \textit{CHIM, a distributed scheme that allocates a random channel to each \text{WBAN}, and provisions backup time-slots for failed transmission. The backup time-slots are scheduled in a way that is similar to DAIL. CHIM enables only a sensor that experiences collisions to hop to an alternative backup channel in its allocated backup time-slot.}
\item \textit{An analytical model that derives bounds on the collision probability and throughput for sensors transmissions.}
\end{itemize}
The simulation results and the mathematical analysis show that our approach can significantly reduce the number of medium access collisions among the individual transmissions of the different coexisting \textit{WBAN}s as well as reduce power consumption. Moreover, \textit{DAIL} and \textit{CHIM} do not require any mutual coordination among the individual \textit{WBAN}s. The organization of the paper is as follows. Section 2 sets our work from other approaches in the literature. Section 3 describes the system model and provides a brief overview of Latin squares. Sections 4 and 5, receptively, describe \textit{DAIL} and \textit{CHIM} in detail. The performance of \textit{DAIL} and \textit{CHIM} are analyzed in 6 and 7, respectively. Section 8 presents the simulations results of \textit{DAIL} and \textit{CHIM}. Finally, the paper is concluded in Section 9.
\section{Related Works}
Co-channel interference avoidance and mitigation have been subject to extensive research in recent years. In the context of \textit{WBAN}s, published techniques can be categorized as resource allocation, power control, some solutions are also based on incorporation of multiple medium access arbitration mechanisms and link adaptation. In the balance of this section, we summarize the major work in each category and compare with our approach. We also discuss previous work that employed Latin squares in medium access arbitration. 
\subsection{Resource Allocation}
Countering interference by careful resource management and medium access scheduling is obviously a viable option. Resource allocation, e.g., channels and time, is an effective way for avoiding radio co-channel interference and medium access collision. However, the main problem in channel assignment solutions is the limited number of available channels, especially, when there is high-density of coexisting \text{WBAN}s. Some approaches have pursued this methodology. Movassaghi et al. \cite{key7, key8} proposed an adaptive inter-\textit{WBAN} interference mitigation scheme, where sensors transmissions are classified according to the QoS parameters. In their scheme, each interfering pair of \textit{WBANs} generates a table of interfering sensors and determines an interference region (IR) between them. An orthogonal channel is allocated to each sensor in IR that has the highest priority according to QoS classification; while sensors that are not belonging to IR can transmit in the same time period. This scheme reduces the number of assigned channels and improves the throughput. In \cite{key132}, the same authors adapt channel assignment to various mobility scenarios of \textit{WBAN}s using predictive models of their next location. In addition, their proposed algorithm allocates a transmission time such that the interference is better avoided to extend the \textit{WBAN} lifetime and improve the spatial reuse and throughput. Meanwhile, an adaptive scheme based on social interaction is presented in \cite{key1119} to optimize the transmission time allocation. Such scheme factors in the \textit{WBAN} mobility, the density of sensors in a \textit{WBAN}, the received signal strength indicator (RSSI), and the heterogeneous traffic load in order to minimize the power consumption and increase the throughput. 

On the other hand, some approaches tried to avoid interference by assigning conflict free channels. Obviously this approach does not suit dynamic environments where \text{WBAN}s accidently becomes in range of each other. Nonetheless, conflict free channel assignment can fit scenarios where the set of \text{WBAN}s that may coexist can be predicted beforehand. The popular methodology for channel assignment in this case is to use graph coloring. Huang et al. \cite{key22} proposed a distributed scheme that maps the channel allocation as a graph coloring problem. In their scheme, they mapped the time-slots to different colors and further proposed a coloring algorithm to determine a color assignment for each node. The coordinators exchange control messages to achieve a conflict-free coloring. Meantime, Cheng et al. \cite{key174} proposed a random incomplete coloring scheme (RICC) to realize conflict-free channel allocation without involving heavy computation. Although RICC allows for higher spatial reuse of channel allocation, it is still far from the optimal as it only considers the scheduling at the \textit{WBAN} level but not at the node-level, and hence the spatial reuse is not optimally utilized. Similarly, Seo et al. \cite{key1124} presented a coloring-based scheduling method to avoid the inter-\textit{WBAN} interference by allocating different time-slots to adjacent \textit{WBAN}s and by allocating more time-slots to traffic-intensive \textit{WBAN}s to increase the total throughput. Overall, channel and time resource allocation protocols may work efficiently in dynamic environment and under high interference conditions, i.e., highly mobile and densely deployed \text{WBAN}s. These protocols may support an acceptable level of QoS requirements. Graph-based resource allocation protocols are unsuitable for topology with high-frequent changes, e.g., \text{WBAN}s, because of the incurred cost due to update and message exchanges; therefore these protocols may be not only inefficient but also detrimental for health care applications in dynamic environment. Moreover, they do not even support an acceptable level of QoS requirements to sensors. 
\subsection{Power Control}
One of the most design issues in \text{WBAN}s is the scarce energy resource of their sensors. Thus, saving energy by adjusting transmission power of the radio transceiver is very crucial to extend the lifetime of the \text{WBAN}. In \text{WBAN}s, the link quality is affected by different channel conditions such as fading, path loss and shadowing that change frequently due to the body movements. Therefore, the transmission power can be adaptively controlled based on its link state in order to improve the relaibility while minimizing the energy consumption. Kim et al. \cite{key1121} presented a link state estimation transmit power control protocol (TPC) to improve the energy efficiency and the link reliability in \textit{WBAN}s. Their protocol adapts the transmit power according to short-term and long-term link-state estimations of RSSI under different mobility scenarios of the patient. Meantime, Chen et al. \cite{key2054} proposed a power allocation algorithm based on genetic algorithms to mitigate inter-\textit{WBAN} interference and increase the power savings while ensuring QoS guarantees. Yet, their algorithm did not consider the mobility which leads to a long convergence time. In HOSS \cite{key1127}, low interfering sensors transmit on the same channel; while high interfering sensors transmit using orthogonal channels, at the same time, the remaining sensors are associated with an energy harvesting model to gather the wireless broadcast energy from other sensors transmissions. HOSS captures the interference signal to be used as a source of energy and mitigating the interference at the same time. Consequently, the higher the number of coexisting \textit{WBAN}s implies the higher the amount of harvested energy and the higher the network lifetime. 

Zou et al. \cite{key21} proposed a Bayesian non-cooperative game based power control approach to mitigate inter-\textit{WBAN} interference. They modeled \textit{WBAN}s as players and active links as types of players in the Bayesian model to maximize the throughput and energy efficiency of \textit{WBAN}s. Meanwhile, Dong et al. \cite{key1129} proposed a non-cooperative social-based game theoretic transmit power control scheme to maximize the throughput amongst coexisting \textit{WBAN}s so that the average transmission power is minimized. Meantime, Moovasi et al. \cite{key1110} proposed a non-cooperative game-theoretic approach to investigate the problem of joint relay selection and power control in \textit{WBAN}s. To better mitigate the interference and ensure energy efficient communication, each sensor seeks a strategy to select the appropriate transmission power level and the best relay in order to ensure short delay and jitter in a distributed manner. Overall, the published protocols have shown that power control is robust to different body mobility scenarios and suit highly populated environment with \text{WBAN}s. Although link-state based protocols do not require negotiation and message exchange among \text{WBAN}s, they significantly consume energy and provide poor QoS support. These protocols are not recommended for dynamic environments because the link state varies very frequently due to body movements. On the other hand, game-based power control protocols do not support the mobility of \text{WBAN}s. Nonetheless, these protocols support dynamic channel conditions, e.g., varying channel gain, and do not require message exchange, which reduces the energy consumption across coexisting \text{WBAN}s. However, game-based protocols do not support QoS and are characterized by long delays. 
\subsection{Link Adaptation}
Interference, in essence, affects the individual wireless links. Therefore, one way to mitigate the effect of interference is to adjust the link parameters. Link adaptation protocols opt to dynamically match the modulation and coding parameters, e.g., data rate, modulation scheme, etc., of the transmitted signal to the channel conditions, namely, the pathloss, SINR, RSSI, etc. These protocols invariably require some channel state information at the transmitter. For example, the link data rate could change based on the SINR of the channel. Yang et al., \cite{key19} proposed several schemes such as adaptive modulation, adaptive data rates, and duty cycles. The coordinator of a \textit{WBAN} selects the appropriate scheme for use by the sensors based on the level of experienced interference. Whilst, Moungla et al. \cite{key1114} pursued a tree-based topology design for a single mobile \textit{WBAN} to ensure reliable data delivery. \textit{WBAN} sensors share a small number of channels, whereas, the relay nodes share the most number of channels to improve the data flow across the \textit{WBAN}. To mitigate the interference among relays, adaptive data rate and duty cycle are used to improve the data rate and ensure an energy efficient and reliable communication within a \textit{WBAN}.

A relay is defined as an intermediate node connecting a sensor to a coordinator. As stated in the \textit{ZIGBEE} standard \cite{key5a},\cite{key5}, the two-hop communication is a very promising solution as it exploits the spatial diversity to improve \textit{WBAN} energy efficiency and reliability as well as to enable better \textit{WBAN}'s interference mitigation. Dong et al. \cite{key12} proposed a two-hop cooperative scheme where a TPC mechanism is integrated into sensor and relay nodes to prolong their batteries lifetimes and mitigate the interference by increasing the \textit{SINR} of data packets received at the \textit{WBAN} coordinator. Moreover, in \cite{key1126} the same authors implemented a two-hop communication scheme (LRCS) for a \textit{WBAN}. LRCS considered two relay nodes and provides a 3-link diversity gain to the coordinator, wherein selection combining, the coordinator selects the best of the three links according to best link channel gain. Overall, link adaptation protocols do not suite highly mobile and densely deployed \text{WBAN}s because of the fast-changing channel conditions. However, these protocols may introduce some additional energy cost to the relaying nodes and latency to the packet delivery that may be unacceptable in time-sensitive health-care applications. 
\subsection{Use of Multiple Access Protocols}
With contention-based MAC, sensors within a \text{WBAN} can decide their medium access pattern without the need for any time synchronization. However, the individual performance of these sensors may be degraded because of the high overhead incurred due to addressing the medium access collisions when the density of \text{WBAN}s is high; such overhead is mainly due to time and energy consumption during backoffs. On the other hand, contention-free protocols use time synchronization to achieve collision-free transmissions and high throughput. However, one of the major limitations of contention-free approach is the need for time synchronization which is very costly to achieve in \text{WBAN}s. Another limitation is the time-slot allocation, which becomes challenging particularly when the different coexisting \text{WBAN}s employ non-similar duty cycles, i.e., the number of active sensors in a period of time is not consistent among these \text{WBAN}s. Contention-free MAC is reliable and energy-efficient \cite{key1119} in relatively low-density \text{WBAN}s, though extra energy is consumed for their periodic synchronization and control messages.

Chen et al., \cite{key15} proposed a two-layer MAC protocol to deal with the interference of \textit{WBAN}s (2L-MAC). The \textit{WBAN}'s coordinator schedules transmissions within its \textit{WBAN} using TDMA, and a carrier sense mechanism to deal with inter-\textit{WBAN} interference. 2L-MAC reduces packet collisions, transmission delay and energy consumption in sensors. Meantime, Deylami et al. \cite{key1130} proposed a dynamic coexistence management mechanism as an extension to the IEEE 802.15.4 standard \cite{key5} which makes the IEEE 802.15.4-based \textit{WBAN}s able to detect and mitigate the interference. Meanwhile, in the approach of Huang et al. \cite{key1105}, the coordinator employs carrier sense multiple access with collision avoidance (\textit{CSMA/CA}) to adaptively adjust superframe length according to the level of interference, and beacons to control the medium access within its \textit{WBAN}. Similarly, Zhou et al. \cite{key1107} proposed an opportunistic scheduling scheme (OS-MAC) by applying heuristic scheduling to improve the \textit{WBAN} performance. OS-MAC dynamically adjusts the superframe length according to the channel status. On the other hand, Liu et al. \cite{key1111} proposed a TDMA-based MAC protocol (CA-MAC) to maintain an accepted level of QoS while at the same time ensuring energy-efficient communication. CA-MAC dynamically adjusts the transmission order and transmission duration of the sensor nodes according to dynamic channel conditions and traffic characteristics of \textit{WBAN}s under different mobility scenarios.

Grassi et al. \cite{key2026} employed a mechanism that reschedules beacons to avoid active period overlapping and hence reducing the interferences amongst \textit{WBAN}s. The rescheduling may cause significant transmission delay if there is a large number of \textit{WBAN}s. Meantime, Deylami et al. \cite{key1116} proposed a distributed approach that collaboratively arranges the active durations of the superframes of the coexisting \textit{WBAN}s so that interference is minimized and the channel usage is maximized. Similarly, Kim et al. \cite{key2039} proposed a distributed asynchronous inter-\textit{WBAN} interference avoidance scheme (AIIA) that includes the timing offset and dynamically adjust the location of the \textit{TDMA} period to prevent inter-\textit{WBAN} interference by overlapping TDMA transmissions between nearby \textit{WBAN}s. \textit{AIIA} adapts to the level of interference in the mobile \textit{WBAN} environment and improves the coordination time without incurring significant complexity overhead. Such time sharing based solutions in which \textit{WBAN}s interleave their active periods through negotiation or contention are ineffective when the load in \textit{WBAN}s is heavy and duty cycle of \textit{WBAN} is high. Both contention-free and contention-based approaches are suitable for relatively low-density of \text{WBAN}s with low-occupancy channels and few sensors \cite{key2032,key2033,key2024}. However, these approaches are unsuitable for highly mobile \text{WBAN}s with high-density of sensors and with heavy traffic load.
\subsection{Medium Sharing using Latin Square}
Ju et al. \cite{key25} proposed a noteworthy approach although it is not geared for \textit{WBAN}. Basically, they presented a multi-channel topology transparent algorithm that leverages the properties of Latin squares for scheduling the transmissions in a multi-hop packet radio network. Wireless nodes are considered to be equipped with one transmitter and multiple receivers. It is worth noting that in \cite{key10} Latin squares are used in cellular networks for the sub-carrier allocations to users, where a user could be allocated multiple virtual channels. Each virtual channel hops over different sub-carriers at different orthogonal frequency division multiplexing (OFDM) symbol times. Basically, users are allocated multiple sub-carrier-to-OFDM-symbol-time combinations to avoid inter-cell interference. In \textit{DAIL}, sensors are simply allocated a single channel to time-slot combination and this simplifies inter- \textit{WBAN} coordination and time synchronization. In \cite{key300}, multiple topology-dependent transmission scheduling algorithms have been proposed to minimize the \textit{TDMA} frame length in multi-hop packet radio networks using \textit{Galois field theory} and \textit{Latin squares}. For a single-channel network, the \textit{modified Galois field design} and the \textit{Latin square design} for topology-transparent broadcast scheduling is proposed. \textit{Modified Galois field design} obtains much smaller \textit{frame length} than the existing scheme while the \textit{Latin square design} can even achieve possible performance gain when compared with the \textit{modified Galois field design}. In one-hop rather than multi-hop communication scheme, like \textit{DAIL}, using Latin squares yields better schedules the medium access and consequently significantly diminishes the inter-\textit{WBAN} interference. Like \cite{key25}, \textit{DAIL} and \textit{CHIM} exploit the properties of Latin rectangles to generate a predictable interference-free transmission schedule for all sensors within a \textit{WBAN}. However, \textit{DAIL} and \textit{CHIM} depend on the existence of only one rather than multiple receivers within each node and one-hop rather than multi-hop communication. Unlike \textit{DAIL}, \textit{CHIM} minimizes the frequency of channel switching significantly, i.e., \textit{CHIM} applies channel switching only when a sensor experiences interference. 

Compared to the related work covered in this section, our approach can be viewed as combing two solution strategies, multi-channel and time-slot adjustment. In this paper, we employ Latin rectangles for interference mitigation amongst multiple coexisting \textit{WBAN}s. To mitigate interference, our approach opts to exploit the availability of multiple channels in the \textit{ZIGBEE} standard and employs Latin rectangles as the underlying scheme for channel and time-slot allocation to sensors while enabling autonomous scheduling of the medium access. By leveraging the properties of Latin rectangles, our approach reduces the probability of medium access collision and provides better usage of the limited resources of \textit{WBAN}s. Since the potential for medium access collision grows with the increase in the communication range and the density of sensors in the individual \textit{WBANs}, we propose two schemes. The first scheme, namely, \textit{DAIL} suits crowded areas where a high density of \textit{WBAN}s coexists. On the other hand, the second scheme, \textit{CHIM}, takes advantage of the relatively lower density of coexisting \textit{WBAN}s to save the power resource at both sensor- and \textit{WBAN}-levels.
\section{System Model and Preliminaries}
\subsection{System and Problem Models}
We assume \textit{N} TDMA-based \textit{WBAN}s may coexist in the close proximity of each other. As an example, when a number of patients move around in a hospital's hall or corridor. Each \textit{WBAN} is formed of one coordinator denoted by \textit{Crd} and up to \textit{K} wireless sensors. Each sensor generates its sensed data according to a predetermined sampling rate and transmits at a maximum date rate of 250Kb/s using the unlicensed \textit{2.4 GHz} band. In our approach, we assume that all \textit{ZIGBEE} standard channels are available for sensors and the coordinator within each \textit{WBAN} at any time. Furthermore, all coordinators have richer power supply than sensors and are not affected by the frequent channel hopping.

\begin{figure}
  \centering
       \includegraphics[width=0.275\textwidth,height=0.2\textheight]{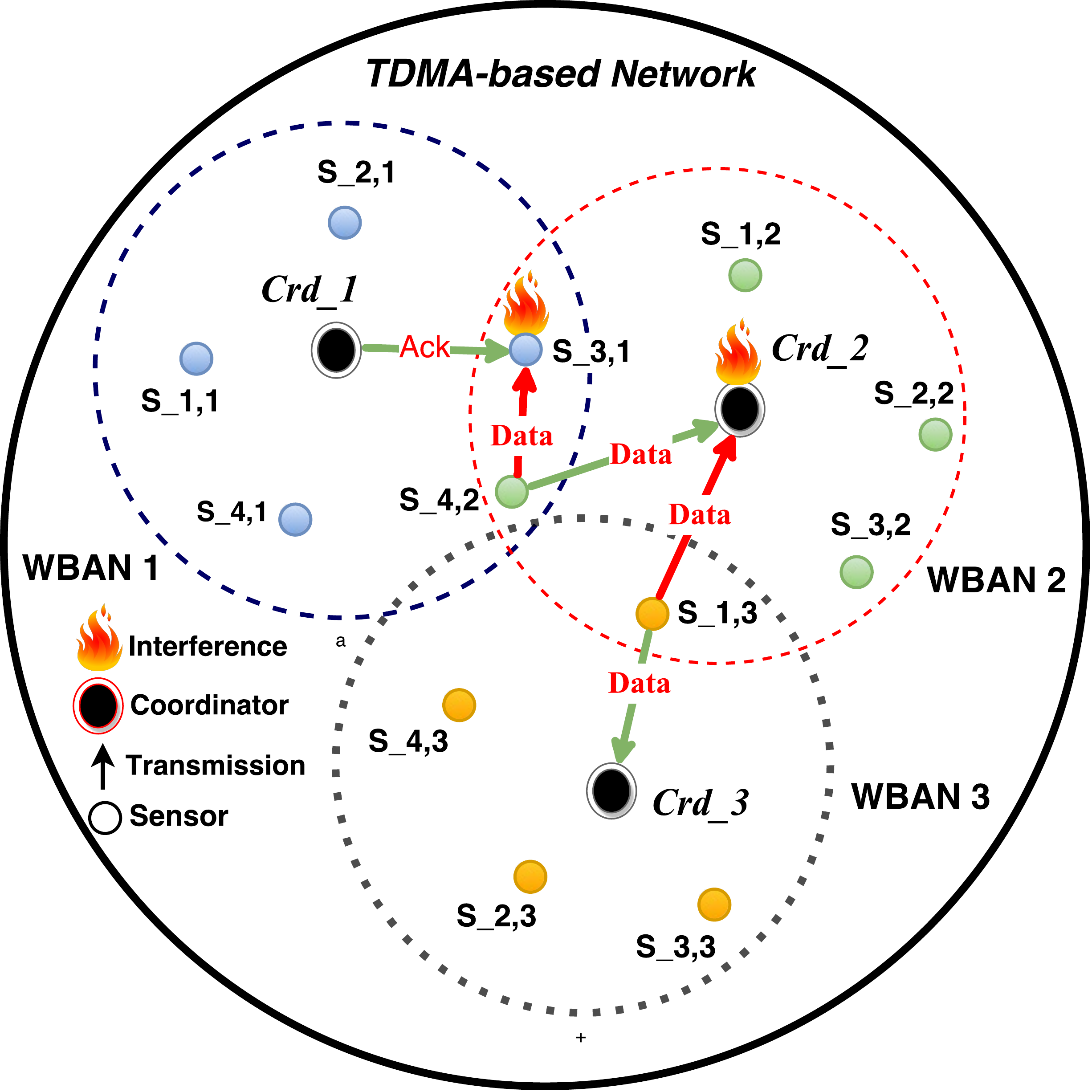}
\caption{Collision scenarios at sensor- and coordinator-levels}
\label{collision}
\end{figure}

Due to the \textit{WBAN}'s unpredictable motion pattern, it is very hard to achieve inter-\textit{WBAN}s coordination or to have a central unit to mitigate the potential interference when some of them are in close proximity of each other. Basically, the co-channel interference happens when multiple \textit{WBAN}s concurrently access the same channel and their corresponding radio transmission ranges overlap. In essence, the medium access collision may arise due to the collisions amongst the concurrent transmissions of these \textit{WBAN}s. Since data packets may be lost due to the co-channel interference, and hence acknowledgments are required to confirm successful reception. Time-outs are used to detect reception failure at the corresponding receivers. We note that collisions may take place at the level of data or acknowledgement packets as shown in \textbf{Figure \ref{collision}} and explained below. \textbf{Table \ref{symbol}} shows symbols descriptions.
\begin{table}
\centering
\caption{Symbol description}
\label{symbol}
\resizebox{0.3\textwidth}{!}{
\begin{tabular}{ll}
\noalign{\smallskip}\hline
\hline\noalign{\smallskip}
$WBAN_k$&$k^{th}$ \textit{WBAN}s\\
$Crd_q$&coordinator of $q^{th}$ \textit{WBAN}s\\
$S_{i,k}$&$i^{th}$ sensor of $k^{th}$ \textit{WBAN}s\\
$C_k$ &a default channel of $k^{th}$ \textit{WBAN}s\\
$BKC(S_{i,k})$&a backup channel picked by $i^{th}$\\
&sensor of $k^{th}$ \textit{WBAN}s\\
\noalign{\smallskip}\hline
\hline\noalign{\smallskip}
\end{tabular}}
\end{table}
\subsubsection{Data Packets Collision}
Essentially, a sensor's data packet experiences a collision at the receiving coordinator when this latter is in the transmission or radio range of another active sensor or coordinator. When a sensor denoted by $S_{i,k}$ of $\textit{WBAN}_k$ transmits to its corresponding coordinator denoted by $Crd_k$ while another sensor $S_{j,q}$ or coordinator $Crd_q$ of another $\textit{WBAN}_q$ transmits using the same channel that $S_{i,k}$ employs, i.e., a collision occurs under the following condition: $Crd_k$ is in range of $Crd_q$ or $S_{j,q}$; and $Crd_q$ or $S_{j,q}$ transmits using the same channel used by sensor $S_{i,k}$. In essence, such collision may be experienced in both \textit{DAIL} and \textit{CHIM} schemes, when $C_k = C_q$; i.e., both $\textit{WBAN}_k$ and $\textit{WBAN}_q$ happen to pick the same channel for intra-\textit{WBAN} communication, in which case $S_{j,q}$ or $Crd_q$ could be sending a data or acknowledgement packets, respectively. In addition, another collision scenario may be experienced only in the case of \textit{CHIM} scheme when $C_k$ = $BKC(S_{j,q})$ or $C_q = BKC(S_{i,k})$; i.e., the channel has been picked by $\textit{WBAN}_k$ is equal to the same channel that has been allocated to sensor $S_{j,q}$ of $\textit{WBAN}_q$ in its backup time-slot within the \textit{IMB} frame which will be defined in section 5-A. In \textit{IMB} frame, each interfering sensor may use a backup channel within its allocated backup time-slot.    
\subsubsection{Acknowledgment Packets Collision}
In similar way, a coordinator's acknowledgment packet experiences a collision at the receiving sensor when this latter is in the transmission or radio range of another active sensor or coordinator. When $S_{i,k}$ receives from its corresponding $Crd_k$, while at the same time, another $S_{j,q}$ or $Crd_q$ transmits using the same channel that $S_{i,k}$ uses, i.e., a collision occurs under the following condition: $S_{i,k}$ is in transmission range of $Crd_q$ or $S_{j,q}$; and $Crd_q$ or $S_{j,q}$ transmits using the same channel employed by $S_{i,k}$.
\subsection{Latin Squares}
In our approach, we have employed the special properties of Latin squares to allocate interference mitigation channels. In this section, we provide a brief overview of Latin squares.
\begin{definition}
\textit{A Latin square is a $K \times  K$ matrix, filled with \textit{K} distinct symbols, each symbol appearing once in each column and once in each row.}
\end{definition}
\begin{definition}\label{orthogonal}
\textit{Two distinct $K \times  K$ Latin squares \textit{E = ($e_{i,j}$)} and \textit{F = ($f_{i,j}$)}, so that $e_{i,j}$ and $f_{i,j}$ $\in$ $\{1,2, \dots K\}$, are said to be orthogonal, if the $K^{2}$ ordered pairs ($e_{i,j},f_{i,j}$) are all different from each other. More generally, the set \textit{$O=\{E_{1}, E_{2}, E_{3},\dots, E_{r}\}$} of distinct Latin squares \textit{E} is said to be orthogonal, if every pair in \textit{O} is orthogonal.}
\end{definition}
\begin{definition}
\textit{An orthogonal set of Latin squares of order \textit{K} is of size \textit{(K-1)}, i.e., the number of Latin squares in the orthogonal family is \textit{(K-1)}, is called a complete set \cite{key25}.}
\end{definition}
\begin{definition}
\textit{A $M \times  K$ Latin rectangle is a $M \times  K$ matrix \textit{G}, filled with symbols $a_{ij}$ $\in$ $\{1,2,\dots,K\}$, such that each row and each column contains only distinct symbols.}
\end{definition}
\begin{theorem}\label{theo1}
If there is an orthogonal family of \textit{r} Latin squares of order \textit{K}, then \textit{$r\leq K-1$} \cite{key26}
\end{theorem}
\textit{E} and \textit{F} are orthogonal Latin squares of order \textit{3}, because no two ordered pairs within \textit{E $\bowtie$ F} are similar.
\begingroup\makeatletter\def\f@size{8}\check@mathfonts
\begin{center}
$E = 
\begin{bmatrix}
   1&2&3 \\
   2&3&1 \\
   3&1&2 \\
\end{bmatrix}$ $F = 
\begin{bmatrix}
   1&2&3 \\
   3&1&2 \\
   2&3&1 \\
\end{bmatrix}$
$E\bowtie F = 
\begin{bmatrix}
   1,1&2,2&3,3\\
   2,3&3,1&1,2\\
   3,2&1,3&2,1\\
\end{bmatrix}$
\end{center}
\endgroup
According to \textit{\textbf{Theorem \ref{theo1}}}, the number of \textit{WBAN}s using orthogonal Latin squares is upper bounded by \textit{K-1}, thus, \textit{K} should be large enough so that, each \textit{WBAN} can pick an orthogonal Latin square with high probability. According to the orthogonality property, each \textit{WBAN} will be assigned a unique channel allocation pattern that does not look like the pattern of other \textit{WBAN}s to avoid the co-channel interference, i.e., they do not share the same symbol positions, each in its own Latin rectangle and consequently, no other \textit{WBAN} in the network would simultaneously share the same pattern with $\textit{WBAN}_i$ all the time. Generally, our approach makes it highly improbable for two transmissions to collide. Nonetheless, a medium access collision may still happen when (i) two or more coexisting \textit{WBAN}s in the proximity of each other randomly pick the same Latin rectangle, or (ii) when the number of \textit{WBAN}s coexisting in the proximity of each other is larger than the number of \textit{ZIGBEE} standard channels forming the Latin rectangle which is \textit{16} in our case.

Basically, a particular \textit{WBAN} randomly picks one Latin square from the orthogonal set of Latin squares. According to their properties, Latin squares ensure that a particular \textit{WBAN} will not share a channel amongst the coexisting \textit{WBAN}s that use distinct Latin squares chosen from the same orthogonal set. The Latin square size will be determined according to the largest between the number of sensors that forms each \textit{WBAN}, denoted by \textit{K} and the number of channels that determines the number of the rows in the Latin, denoted by \textit{M}. In the IEEE 82.15.6 standard \cite{key5}, the number of channels is limited to \textit{16}, i.e., the maximum number of rows in the Latin is \textit{16}. Therefore, the possible number of simultaneous transmissions that can be scheduled is no more than \textit{16}. Due to the limited number of possible channels, \textit{DAIL} addresses this limitation by employing Latin rectangles rather than Latin squares. In other words, \textit{DAIL} does not fix the number of columns in the Latin square, i.e., the value of \textit{K} is not fixed to \textit{16}. Hence \textit{DAIL} supports \textit{$K > M$}, i.e., more than \textit{16} transmissions can be scheduled. Unlike \textit{DAIL}, \textit{CHIM} optimizes the time-slot allocation within the superframe and the channel hopping is decided only when the interference is experienced.
\section{Distributed Interference Avoidance using Latin Rectangles Scheme (\textit{DAIL})}
\subsection{Algorithm Description}
\textit{DAIL} operates in a distributed manner and suits crowded area with a large number of patients, each may carry a single \textit{WBAN} and may move towards other patients, e.g., in a hospital's corridor. Through using Latin rectangles, \textit{DAIL} assigns a channel and time-slot combinations that reduce the probability of medium access collision. In \textit{DAIL}, each coordinator randomly selects one orthogonal Latin rectangle through which it assigns a unique symbol to each sensor within its \textit{WBAN}. According to its assigned symbol, a sensor determines its transmission schedule which is formed of a sequence of channel and time-slot combinations. That means each sensor determines its hopping pattern, i.e., which hopping channel to use in which time-slot within every superframe.

\textit{DAIL} enables different coexisting sensors to hop among distinct channels to avoid the collision among their corresponding transmissions that happen in the same time-slot. Thus, the number of collisions depends on the number of coexisting \textit{WBAN}s, i.e., the corresponding interfering sensors and the number of orthogonal Latin rectangles used by the interfering \textit{WBAN}s. Therefore, the medium access collision among the transmissions of the different sensors is completely avoided, iff, the number of orthogonal Latin rectangles is larger than the number of those sensors competing to transmit in the same time-slot. Otherwise, \textit{DAIL} extends the number of columns in the Latin rectangle which is directly related to the length of the \textit{WBAN}'s superframe by adding extra time-slots to lower the probability of medium access collisions. For example, if the number of coexisting \textit{WBAN}s is \textit{N} and the number of Latin rectangles is \textit{P}, each of size \textit{16 $\times$ K}, where \textit{K} denotes the number of columns in the Latin and at the same time the number of time-slots in the superframe. If \textit{N $>$ max(16,K)}, then each \textit{WBAN} will extend the number of columns in the Latin, i.e., the number of time-slots in the superframe from \textit{K} to \textit{K + XT}, where \textit{XT = N - max(16,K)}. Doing so, such sensors will have a higher probability to not pick the same channel in the same time-slot and hence the number of collisions is minimized. \textbf{Algorithm \ref{dail}} provides a high level summary of \textit{DAIL}.
\setlength{\textfloatsep}{1pt}
\begin{algorithm}
\footnotesize
\SetKwData{Left}{left}\SetKwData{This}{this}\SetKwData{Up}{up}
\SetKwFunction{Union}{Union}\SetKwFunction{FindCompress}{FindCompress}
\SetKwInOut{Input}{input}\SetKwInOut{Output}{output}

\Input{\textit{N} \textit{WBAN}s, \textit{K} sensors/\textit{WBAN}, Coordinator \textit{Crd}, \textit{M} \textit{ZIGBEE} channels, Latin rectangle \textit{R}, frame length \textit{FL}}

\textbf{BEGIN}
 
\quad \textit{FL =  K}\: //\; default setting of the frame length 
 
\quad \textit{\textbf{if}} \textit{N $>$ K} \textit{\textbf{then}}

\quad \quad \quad \textit{FL = N}\: //\; \textit{Crd} increases the number of time-slots in the superframe

\quad\textit{\textbf{endif}}

\quad Each \textit{WBAN}'s \textit{Crd} randomly picks a Latin rectangle \textit{R} of size \textit{M $\times$ FL}

\textbf{END}

\caption{Proposed \textit{DAIL} Scheme}
\label{dail}
\end{algorithm}
\DecMargin{1em}
\subsection{\textit{DAIL} Superframe Structure}
\begin{figure}
  \centering
       \includegraphics[width=0.275\textwidth,height=0.125\textheight]{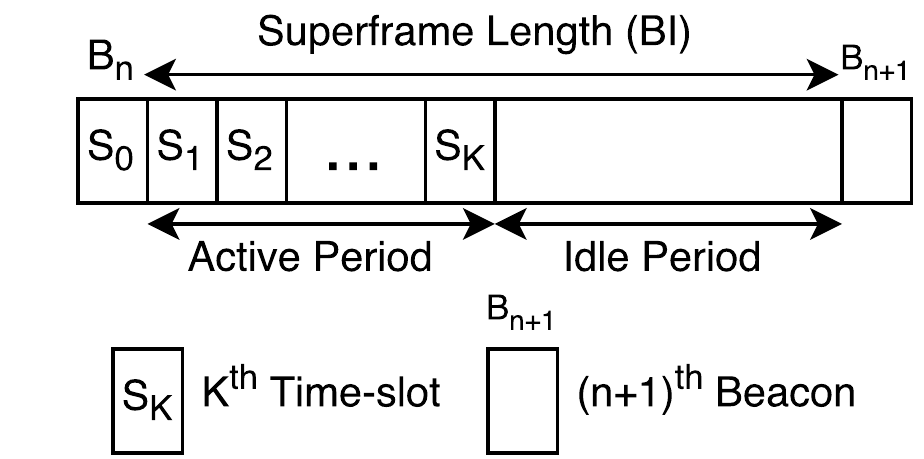}
\caption{Proposed superframe structure for \textit{DAIL}}
\label{superframestructdial}
\end{figure}

Each \textit{WBAN}'s superframe is delimited by two beacons and composed of two successive frames: (i) the active frame which is dedicated for all sensors within a \textit{WBAN} and, (ii) the inactive frame which is designated for the \textit{WBAN}'s coordinator as shown in \textbf{Figure \ref{superframestructdial}}. Through assuming \textit{M = 16} channels of the \textit{ZIGBEE} are available at each \textit{WBAN}, it is still needed to determine the number of time-slots forming each row of the Latin rectangle and consequently the size of each superframe. In fact, the superframe size depends on how big the time-slot which is based on the protocol in use and on the number of time-slots required which is determined by the different sampling rates of the sensors. Generally, the superframe size is determined based on the highest sampling rate and the sum of the number of samples for all sensors in a time period. \textit{DAIL} requires the same superframe size for all \textit{WBAN}s so that medium access collision could be better avoided by picking the right value for \textit{K}, where \textit{K} is the number of time-slots to be made in the superframe, respectively, in the Latin rectangle.

\subsection{Illustrative Example}
We illustrate our approach through a scenario of 3 coexisting \textit{WBAN}s, where each circumference represents the interference range as shown in \textbf{Figure \ref{collision}}. Furthermore, each \textit{WBAN} is assigned M = 4 channels and consists of L = 4 sensors, in turn, each sensor is assigned a symbol from the set K = \{1,2,3,4\}$\iff$\{G,B,R,W\}. Here, we assume that each sensor requires only one time-slot to transmit its data in each superframe. Based on this scenario, any pair of sensors are interfering with each other, i.e., they transmit using the same channel at the same time if both sensors are in the intersection of their corresponding interference ranges.

\begin{figure}
  \centering
        \includegraphics[width=0.275\textwidth,height=0.125\textheight]{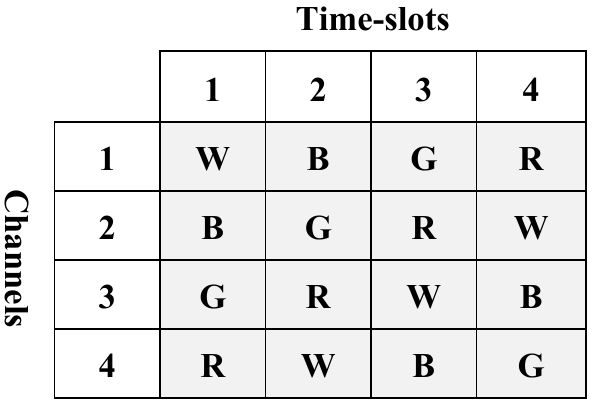}
\caption{A $4 \times 4$ channel to time-slot assignment Latin square}
\label{colors}
\end{figure}

However, as shown in \textbf{Figure \ref{collision}}, $4^{th}$ sensor of $\textit{WBAN}_1$ denoted by $S_{1,4}$ and $S_{2,4}$ are interfering, also, $S_{3,1}$ and $S_{2,3}$. Therefore, in \textit{DAIL}, each \textit{WBAN} picks a distinct Latin rectangle from an orthogonal set as follows: $\textit{WBAN}_1$ picks E, $\textit{WBAN}_2$ picks F and $\textit{WBAN}_3$ picks J, where E and F are considered as in (3-B). Assume 3 sensors, u, v and w of $\textit{WBAN}_1$, $\textit{WBAN}_2$ and $\textit{WBAN}_3$ are, respectively, assigned symbols B, R and G in Latin rectangles E, F and J. Thus, the distinct positions of symbol B in E corresponds to the transmission pattern of u in $\textit{WBAN}_1$'s superframe, similarly for v and w in $\textit{WBAN}_2$ and $\textit{WBAN}_3$, respectively. However, B=2 in E, R=3 in F and G=1 in J, therefore, the transmission patterns for u, v and w are, respectively, represented by B, R, and G of the matrix shown in \textbf{Figure \ref{colors}}. As clearly seen in this figure that u, v and w neither share the same channel nor the same time-slot, i.e., no collision occurs at all. 
\textit{DAIL} is completely distributed channel allocation and medium access scheduling scheme that suits high density of \textit{WBAN}s and does not require coordination among coordinators. \textit{DAIL} assigns \textit{WBAN} sensors channel and time-slot combinations to diminish the probability of inter-\textit{WBAN} interference, i.e., reduces the probability of medium access collision. Nonetheless, \textit{DAIL} drains the power resource of the whole \textit{WBAN}s when some of their corresponding sensors do not face any collision. For example, as an estimate of power cost of a \textit{WBAN} consisting of up to \textit{L} sensors, $\textit{L} < \textit{K}$, is $L\times \textit{HE}$, where \textit{HE} is the power consumption resulting from a channel hopping in each superframe. Therefore, the mean power cost per sensor is denoted $\textit{meanEC} = \frac{L\times HE}{K}$.

More specifically, \textit{DAIL} imposes on each \textit{WBAN}'s sensor to hop among the available channels whether that sensor experiences collision or not. Although the interference-free sensors do not need to hop among the channels, and hence the power is wasted at both the sensor- and the \textit{WBAN-} levels. Another shortcoming in \textit{DAIL} is that no more than 16 transmissions can be scheduled and thus, this limits the number of transmitting sensors, i.e., if more than 16 \textit{WBAN}s coexist, and then the collisions may arise. To overcome such issues in \textit{DAIL}, we propose a distributed scheme, namely, \textit{CHIM}, to reduce the number of medium access collisions and overhead at the coordinator- and sensor-levels as well as to save power of the relatively low density of coexisting \textit{WBAN}s.
\section{Channel Hopping for Interference Mitigation Scheme (\textit{CHIM})}
Like \textit{DAIL}, \textit{CHIM} leverages the special properties of Latin squares to enable predictable channel hopping in order to avoid the co-channel interference amongst coexisting \textit{WBAN}s. However, \textit{CHIM} takes advantage of the relatively lower density of interfering \textit{WBAN}s to save the power resource at both sensors- and \textit{WBAN}-levels. Basically, \textit{CHIM} enables only sensors that experience collisions to hop among backup channels, each in its allocated backup time-slot. \textit{CHIM} imposes less overhead at the sensor- and coordinator-levels because of only sensors that experience collisions are required to use their pre-computed transmission schedules (i.e., a combination of a backup channel and a time-slot). As pointed out, the co-channel interference happens if the concurrent transmissions of sensor nodes and coordinator nodes in the different coexisting \textit{WBAN}s collide. To mitigate the potential co-channel interference, \textit{CHIM} exploits the availability of multiple channels in the \textit{ZIGBEE} standard to allocate each \textit{WBAN} a default channel and in the case of interference, it enables the individual interfering sensors to hop among the remaining channels in a pattern that is predictable within each \textit{WBAN} and random to the other \textit{WBAN}s. To do that, \textit{CHIM} modifies the superframe structure and extends its size by adding extra interference mitigation backup time-slots and employs the properties of the Latin rectangles for channel allocation to the sensors within each individual \textit{WBAN}.
\subsection{\textit{CHIM} Superframe Structure}
\begin{figure}
  \centering
       \includegraphics[width=0.3\textwidth,height=0.125\textheight]{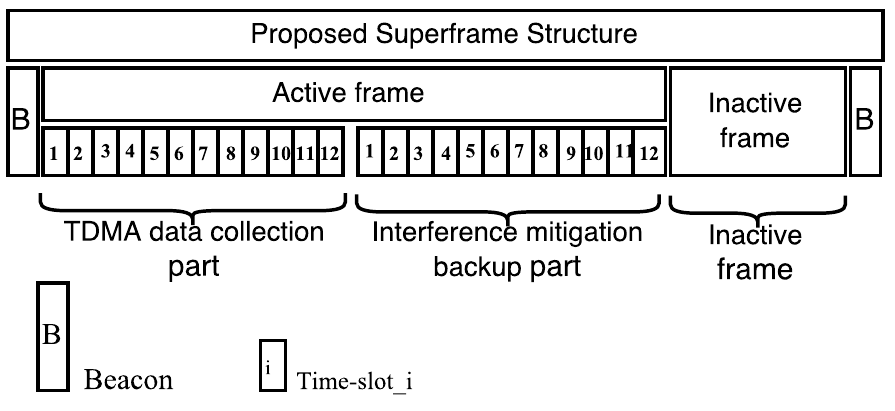}
\caption{Proposed superframe structure for \textit{CHIM}}
\label{superframestructchim}
\end{figure}

Like \textit{DAIL}, in \textit{CHIM}, each \textit{WBAN}'s superframe is composed of successive active and inactive frames as shown in \textbf{Figure \ref{superframestructchim}}, however the active frame is further divided into two parts, the \textit{TDMA} data-collection part and the interference mitigation backup (\textit{IMB}) interference mitigation part where each part consists of \textit{K} time-slots. \textbf{Figure \ref{superframestructchim}} shows \textit{CHIM} superframe structure.
In the \textit{TDMA} data-collection part, each sensor transmits the sensed data packet to the coordinator in its assigned time-slot using the default channel. However, in the \textit{IMB} part, each interfering sensor retransmits the same data packet to the coordinator in its allocated backup time-slot using the priori-agreed upon the channel. In interference-free conditions, the \textit{WBAN}'s coordinator remains tuned to the default operation channel. If communication with a particular sensor $S_{i}$ fails during its designated time-slot, the coordinator will tune to the $S_{i}$'s backup channel during $S_{i}$'s time-slot in the \textit{IMB} part of the active frame of the superframe. However, during the inactive frame, all the active sensors within each \textit{WBAN} turn off their radio transceiver off and sleep, whilst, the corresponding coordinator transmits all the data collected to a command center.

Yet, it is still required to determine the length of each frame. Like \textit{DAIL}, the length of the \textit{TDMA} part is determined according to the largest sampling rate and the sum of the number of samples for all sensors within a \textit{WBAN} in a time period. However, \textit{CHIM} requires the length of the \textit{TDMA} part for all \textit{WBAN}s to be of the same size so that the medium access collision could be better in the backup part. Thus, \textit{CHIM} states that \textit{2 $\times $ K} time-slots should be made in the active frame, i.e., \textit{K} time-slots forms the \textit{TDMA} part and \textit{K} time-slots forms the \textit{IMB} backup part. Whilst, the inactive frame directly follows the active frame and whose length depends on the duty cycle scheme used by the individual sensors within each \textit{WBAN}.
\subsection{Network Setup}
At the setup of the network, each coordinator of a \textit{WBAN} randomly selects a single \textit{$M \times K$} Latin rectangle from the set of the orthogonal Latin rectangles and one \textit{default channel}. Initially, the coordinator commands all the individual sensors within its \textit{WBAN} to utilize the common \textit{default channel} along the \textit{TDMA} data collection part. In addition, each \textit{WBAN}'s coordinator assigns a single \textit{symbol} to each sensor within its \textit{WBAN}. The selected \textit{symbol} from the set \textit{\{{1,2,\dots, K}\}} determines the transmission schedule for each sensor. In other words, the \textit{symbol} reflects position hopping in the rectangle relates one \textit{interference mitigation channel} and one \textit{backup time-slot}. Thereby, the \textit{WBAN}'s coordinator forms the sequence of an \textit{interference mitigation channel} and one \textit{backup time-slot} combinations for each sensor within its \textit{WBAN}. In addition, each \textit{WBAN}'s coordinator reports the sequence of combinations or the transmission schedule corresponding to each sensor within its \textit{WBAN} through beacons. Subsequently, each sensor will be informed about its allocated \textit{interference mitigation backup channel} and \textit{backup time-slot} to be used within the \textit{IMB} part of the superframe.
\subsection{Network Operation under \textit{CHIM}}
\textit{CHIM} relies on the acknowledgment packets and the time-outs to detect the packet collision that occurs at both the \textit{coordinators} and \textit{sensors}. In the superframe's \textit{TDMA} active part, each individual sensor transmits its packet to the coordinator in its allocated time-slot using the default channel. Immediately after this transmission, that sensor sets a timer and then waits for the acknowledgment. If the sensor receives the acknowledgment packet from the \textit{WBAN}'s coordinator, the transmission is considered successful, and then the sensor turns its radio transceiver off and sleeps until the next superframe. Therefore, all the sensors that succeed in their transmissions will not switch to new channels and will not occupy backup time-slots in the \textit{IMB} part. 

However, if the sensor does not receive the acknowledgment packet before the timer expires, the transmission is considered unsuccessful due to the collision and subsequently, it should use its allocated backup channel and time-slot to mitigate the interference. Basically, the sensor that experiences interference should wait until the \textit{TDMA} part completes and then it switches its current default channel to the allocated interference mitigation channel at the beginning of its backup time-slot through which it retransmits the data packet to its corresponding coordinator. In fact, the packet delivery failure results from the data or acknowledgment packets collisions at both the \textit{sensor-} or \textit{coordinator-}levels, respectively. That means, the desired data packet is lost at the coordinator due to the interference that the transmitting sensor experienced from the simultaneous transmissions of sensors or coordinators in other coexisting \textit{WBAN}s; or the acknowledgment packet is lost at the desired sensor due to the interference that the desired sensor experienced from the simultaneous transmissions of the sensors or coordinators in other coexisting \textit{WBAN}s sharing the same channel with the desired sensor. Therefore, depending on the timeouts and the acknowledgment packets, the interfering nodes, i.e., sensors and coordinators, resolve the medium access collision issue in the same way. \textbf{Algorithm \ref{chim}} summarizes the proposed \textit{CHIM} scheme.
\setlength{\textfloatsep}{1pt}
\begin{algorithm}
\footnotesize
\SetKwData{Left}{left}\SetKwData{This}{this}\SetKwData{Up}{up}
\SetKwFunction{Union}{Union}\SetKwFunction{FindCompress}{FindCompress}
\SetKwInOut{Input}{input}\SetKwInOut{Output}{output}

\Input{\textit{N} \textit{WBAN}s, \textit{K} Sensors/\textit{WBAN}, Orthogonal Latin rectangle \textit{OLR}}

\textit{Stage 1: Network Setup}

          \; \textbf{for} i = 1 \textbf{to} N
          
          \; \qquad $Crd_i$ randomly picks a single $DFC_i$ $\&$ $OLR_i$ for its $\textit{WBAN}_i$
          
          \; \qquad \textbf{for} k = 1 \textbf{to} K
          
          \; \qquad \qquad $Crd_i$ allocates $BKC_{k,i}$ $\&$ $BKTS_{k,i}$ to $S_{k,i}$ from $OLR_i$

\textit{Stage 2: Sensor-level Interference Mitigation}

             \; \textbf{for} i = 1 \textbf{to} N
            
             \; \qquad \textbf{for} k = 1 \textbf{to} K
              
             \; \qquad \qquad $S_{k,i}$ transmits $Pkt_{k,i}$ in $TS_{k,i}$ to $Crd_{i}$ on $DFC_i$ in \textit{$TDMA_i$}
            
             \; \qquad \qquad \textbf{if} $Ack_{k,i}$ is successfully received by $S_{k,i}$ on $DFC_i$
            
             \; \qquad \qquad \quad $S_{k,i}$ switches to SLEEP mode until the next superframe
            
             \; \qquad \qquad \textbf{\textit{else}}
            
             \; \qquad \qquad \quad $S_{k,i}$ waits its designated $BKTS_{k,i}$ within $IMB_{i}$ part 
             
             \; \qquad \qquad \quad $S_{k,i}$ retransmits $Pkt_{k,i}$ in $BKTS_{k,i}$ to $Crd_{i}$ on $BKC_{k,i}$
                 
\textit{Stage 3: Coordinator-level Interference Mitigation}     

             \; \textbf{for} i = 1 \textbf{to} N
            
             \; \qquad \textbf{for} k = 1 \textbf{to} K
            
             \; \qquad \qquad \textbf{\textit{if}} $Crd_{i}$ successfully received $Pkt_{k,i}$ in $TS_{k,i}$ on $DFC_i$
            
             \; \qquad \qquad \quad  $Crd_{i}$ transmits $Ack_{k,i}$ in $TS_{k,i}$ to $S_{k,i}$ on $DFC_i$
            
             \; \qquad \qquad \textbf{\textit{else}}
            
             \; \qquad \qquad \quad $Crd_{i}$ will tune to $BKC_{k,i}$ to receive from $S_{k,i}$ in $IMB_{i}$
             
             \; \qquad \qquad \quad $Crd_{i}$ receives $Pkt_{k,i}$ in $S_{k,i}$'s $BKTS_{k,i}$ on $BKC_{k,i}$
                       
\caption{Proposed \textit{CHIM} Scheme}
\label{chim}
\end{algorithm}
\DecMargin{1em}
\section{DAIL Analysis}
We analyze the performance of \textit{DAIL} mathematically and consider a multi-channel TDMA-based\text{WBAN}s.  The superframes of each individual \textit{WBAN} are constructed as from one $M \times K$ matrix. Within the superframe, each sensor may be assigned \textit{\textit{M}} time-slots to transmit its data packet according to a unique channel to time-slot assignment pattern. These patterns are generated from the orthogonal family of $M \times K$ Latin rectangles. However, all the individual sensors of a \textit{WBAN} share one common $M \times K$ Latin rectangle, where the channel to time-slot assignment pattern of each sensor corresponds to a single symbol pattern in the Latin rectangle, as shown in \textbf{Figure \ref{colors}}. 
\subsection{Interference Bound}
In this subsection, we determine the worst-case collision pattern for the individual sensor.
\begin{definition}
\textit{Let E and F be two orthogonal $M \times K$ Latin rectangles. Symbol e from E is assigned to sensor u, and symbol f from F is assigned to sensor v. Then, there exists a collision at the $j^{th}$ slot on $i^{th}$ channel for u and v, if the ordering (e,f) of both rectangles appears at $i^{th}$ row, $j^{th}$ column, which means $[E_{i,j}] = e$ and $[F_{i,j}] = f$.}
\end{definition}
\begin{theorem}\label{coll}
\textit{If two sensors are assigned two distinct symbols in the same Latin rectangle; there will be no collision among their transmissions. If they are assigned symbols from two distinct orthogonal Latin rectangles, then, they will face at most one collision in every superframe.}
\end{theorem}
\textit{\textbf{Proof:} From the definition of Latin rectangles, because every symbol occurs exactly one time in each row and exactly one time in each column, any two time-slot assignment patterns constructed from the same Latin rectangle will not have any overlap in their patterns and so they will not have any collision with each other. Based on \textbf{Definition \ref{orthogonal}}, hence, the ordering (e,f) for any pair of orthogonal Latin rectangles, where, e and f $\in$ \{1,2,\dots, K\}, can only appear one time, which means that these sensors will only have one opportunity of collision.}
\begin{theorem}
In a network of N \textit{WBAN}s, each sensor has a channel to time-slot transmission pattern corresponding to a symbol pattern chosen from one of the $K^{th}$ set of orthogonal Latin rectangles. Let us consider a sensor denoted by \say{S} surrounded by maximum number of \textit{O} \textit{WBAN}s, i.e., \textit{O} sensors from other \textit{WBAN}s, which means, \textit{O} sensors may coexist in the communication range of S. Then, S may experience at most \textit{O} collisions. Additionally, sensor S may face a minimal number of collisions which equal to max(\textit{O}-K+1,0).
\end{theorem}

\textit{\textbf{Proof:} Based on \textbf{Theorem \ref{coll}}, each neighboring sensor can create at most one collision to S. In the worst case, all \textit{O} sensors are within the range of communication of S. The transmissions patterns of \textit{O} sensors are constructed from Latin rectangles that are different from the Latin rectangle utilized by S. Subsequently, the maximum number of possible collisions experienced by S is \textit{O}. Now, to count the minimal number of collisions for S, it is required to find the maximum number of sensors that construct their transmission patterns from the same Latin rectangle, which is K, i.e., K sensors will have no collision according to \textbf{Theorem \ref{coll}}. Also, \textbf{Theorem \ref{coll}} proves that there exists at most one collision for each pair of sensors constructing their transmission patterns from two different orthogonal Latin rectangles. Therefore, each of the remaining sensors (\textit{O}-K+1) will cause one collision to S because they belong to different orthogonal Latin rectangles. As a result, the minimum number of collisions for sensor S surrounded by \textit{O} sensors is equal to max((\textit{O}-K+1),0).}
\subsection{Collision Probability}
We consider a sensor $S_i$ of $\textit{WBAN}_i$ is surrounded by \textit{O} interfering sensors $v_j$ of different coexisting $\textit{WBAN}_j$ in the vicinity, where $j=1,2,\dots,\textit{O}$ and $i\neq j$. For simplicity, we assume, each sensor transmits one data packet in each time-slot. However, sensor $S_i$ successfully transmits its data packet in time-slot t, on channel h to the coordinator, if, none of the \textit{O} neighbors transmits its data packet using the same time-slot on the same channel as sensor $S_i$. We represent the number of sensors that transmit their packets in the same time-slot as sensor $S_i$ as a random variable denoted by X. When x packets are transmitted in the same time-slot as $S_i$, then, the probability of event \textit{X=x} is denoted by \textit{Pr(X=x)} and defined by defined by \textbf{eq.} \ref{eq1} below.

\small 
\begin{equation}\label{eq1}
      \begin{split}
         Pr\left(X=x\right)&=C_{x}^{\textit{O}+1}\times  \omega^{x}\times  (1-\omega)^{\textit{O}-x}\\ 
         &\times \left(min(M,K)/K\right)^{x} \forall\: x\: \leq\: \textit{O}
      \end{split}
\end{equation}
\normalsize
Where $\omega$ is the use factor, defined as the ratio of the time that a sensor is in use to the total time that it could be in use. Now, suppose Y sensors out of X sensors schedule their transmissions according to the same Latin rectangle as sensor $S_i$, i.e. y out of x sensors select symbol patterns from the same Latin rectangle as $S_i$.

\small 
\begin{equation}\label{eq2}
     \begin{split}
        Pr\left(Y=y \mid X=x\right)&=\left(C_{y}^{K+1}\times  C_{x-y}^{Z-K}\right)/C_{x}^{Z-1}\\
         &\forall x \leq \textit{O} \: \& \:  \forall \: y\: \leq \: x
\end{split}
\end{equation}
\normalsize
Where $Z = K \times m$ is the total number of symbol patterns in the orthogonal Latin rectangles family. However, these Y sensors will not impose any collision with $S_i$'s transmission, since they (Y sensors) use the same Latin rectangle as $S_i$. On the other hand, $X-Y$ sensors may collide with the transmission from sensor $S_i$ to the coordinator on the same channel, then the conditional probability of transmission collision is denoted by (\textit{collTx}) and defined by \textbf{eq.} \ref{eq3} below.

\small
\begin{equation}\label{eq3}
\begin{split}
&Pr(collTx\mid\: Y=y\: \& \: X=x)\\
&=1-Pr(succTx\: \mid\: Y=y\: \& \: X=x)\\
&=1-\left((min(M,K)-1)/min(M,K)\right)^{x-y}
\end{split}
\end{equation}
\normalsize
Where $min(M,K)$ represents the number of transmission time-slots for each sensor in each superframe. Then, the probability of a successful data packet transmission from sensor $S_i$ to the coordinator is denoted by $\lambda$ as follows:

\small 
\begin{equation}\label{eq4}
\begin{split}
\lambda =&\sum_{x=0}^{\textit{O}}\sum_{y=0}^{x}Pr(Y=y,X=x)\times\\
        &Pr(succTx\: \mid Y=y\: \&\: X=x)=\\
        &\sum_{x=0}^{\textit{O}}\sum_{y=0}^{x}Pr(Y=y\: \mid\: X=x)\times  Pr(X=x)\times\\
        &Pr(succTx\: \mid\: Y=y\: \&\: X=x)=\\
        &\sum_{x=0}^{\textit{O}}\sum_{y=0}^{x}(C_{x}^{\textit{O}} C_{y}^{K-1} C_{x-y}^{Z-K})/(C_{x}^{Z-1})\times  \omega^x \times  (1-\omega)^{\textit{O}-x}\times \\
        &\left(min(M,K)/K\right)^{x}\times  \left((min(M,K)-1)/min(M,K)\right)^{x-y} 
\end{split}
\end{equation}
\normalsize 
\subsection{Throughput Analysis}
Let the size of the orthogonal family of $K^{th}$ order Latin squares is \textit{m = K-1} and the transmission pattern of each sensor is determined by one of the \textit{$K^{2}$} distinct symbol patterns in the $K\times  K$ Latin square. When $K > M$, each $K \times K$ Latin square can be cut into $M \times K$ Latin rectangle. To assure that every sensor has unique transmission pattern according to these Latin rectangles, ($K \times m \geq N$) must be satisfied, where N is the number of \textit{WBAN}s. Furthermore, it has been proven in \textbf{Theorem \ref{coll}} that the number of collisions (\textit{\# colls}) in each superframe for any two sensors is either one or zero. Assuming the maximum number of neighbors to S is still \textit{O}, then, each sensor will be assigned $min(M, K)$ transmission time-slots in each superframe denoted by SF. We denote by TS the number of successful transmissions for each sensor, $TS_{min}$ and $TS_{max}$ are the lower and the upper bounds of TS, respectively, when \textbf{eq.} \ref{eq5} holds, every sensor will have its throughput in \textbf{eq.} \ref{eq7} and \textbf{eq.} \ref{eq8} as follows:

\small
\begin{equation}\label{eq5}
K \geq TS_{max} \geq TS \geq TS_{min} > 0 
\end{equation}
\begin{equation}\label{eq6}
TS = min(M,K)-(\#\: colls\: per\: SF)
\end{equation}
\begin{equation}\label{eq7}
TS_{max} =
  \begin{cases}
     K-max(\textit{O}-K+1,0)\:  &\: if\: K\: \leq\: M\\
     M-max(\textit{O}-K+1,0)\:  &\: if\: K\: >\: M\\
  \end{cases}
\end{equation}
\begin{equation}\label{eq8}
 TS_{min} =
  \begin{cases}
   K-\textit{O}\: &\: if\: K\: \leq\: M\\
   M-\textit{O}\: &\: if\: K\: >\: M\\
  \end{cases}
\end{equation}
\normalsize
Therefore, to assure that every sensor has a minimal throughput, K should be greater than \textit{O} when $K \leq M$, or M should be greater than \textit{O} when $K > M$. In order to evaluate the performance of our approach, the best and the lowest throughput, respectively, denoted by $T_{max}$ and $T_{min}$ are defined in \textbf{eq.} \ref{eq9} and \textbf{eq.} \ref{eq10}.
\begin{definition}
$T_{max}$ (resp. $T_{min}$) is defined as the ratio of the maximal (resp. minimal) number of successful transmissions in each SF to its length denoted by FL
\end{definition}

\small 
\begin{equation}\label{eq9}
T_{max}=TS_{max}/FL,\: FL = K
\end{equation}
\begin{equation}\label{eq10}
T_{min}=T_{min}=TS_{min}/FL),\: FL = K
\end{equation}
\normalsize
\begin{theorem}\label{theo2}
For given \textit{O}, N and M, the maximal nonzero upper and lower bounds of throughput T are as follows:
\end{theorem}

\small 
\begin{equation}\label{eq11}
1\: \geq\: T\: \geq\: 1-(\textit{O}/M)\: , if\: K\: \leq\: M
\end{equation}
\begin{equation}\label{eq12}
\begin{split}
&M/max(M,\floor{N/m})\geq T \geq\\
&(M-\textit{O})/max(M,\floor{N/m})\: if\: K>M
\end{split}
\end{equation}
\normalsize
\textbf{Proof:} When $K \leq M$, based on \textbf{eq.} \ref{eq9}, the upper and lower bounds of T are as follows:

\small 
\begin{equation}\label{eq13}
\begin{split}
T_{max}&=TS_{max}/FL=\left(K-max(\textit{O}+1-K,0)\right)/K\\
&=1-\left(max(\textit{O}-K+1,0)/K\right)
\end{split}
\end{equation}
\begin{equation}\label{eq14}
T_{min}=TS_{min}/FL=(K-\textit{O})/K=1-\textit{O}/K
\end{equation}
\normalsize
We can deduce from \textbf{eq.} \ref{eq13} and \textbf{eq.} \ref{eq14} that the upper and lower bounds of T will increase with K. Thus, to ensure the minimal throughput is a positive number and each sensor has a unique transmission pattern, then, this inequality; $\textit{O}<K<\ceil{N/m}$ must be satisfied. Also, we can have, $max(\textit{O}+1-K,0)=0$ and $\ceil{N/m}\leq K\leq M$. Therefore, when K = M, the maximal upper and lower bounds of the throughput are shown in \textbf{eq.} \ref{eq15} and \textbf{eq.} \ref{eq16} below.

\small 
\begin{equation}\label{eq15}
T_{max} = 1 \: and \: T_{min} = 1-\textit{O}/M
\end{equation}
\begin{equation}\label{eq16}
T_{min} = 1-\textit{O}/M
\end{equation}
\normalsize
Similarly, if $K>M$, the bounds of T are shown in \textbf{eq.} \ref{eq17} and \textbf{eq.} \ref{eq18} below.

\small
\begin{equation}\label{eq17}
%\begin{split}
 T_{max}=TS_{max}/FL=\left(M-max(\textit{O}+1-K,0)\right)/K=M/K
% \end{split}
\end{equation}
\begin{equation}\label{eq18}
 T_{min}=TS_{min}/FL=(M-\textit{O})/K
\end{equation}
\normalsize
However, these bounds decrease when $K$ increases. So, when $K > \ceil{N/m}$ and $K > M$ are combined, then, $K > max (M, \ceil{N/m})$ is true, and so the maximal upper and lower bounds of T are as in \textbf{eq.} \ref{eq19} and \textbf{eq.} \ref{eq20} below. 

\small 
\begin{equation}\label{eq19}
 T_{max}=M/max(M,\ceil{N/m})
\end{equation}
\begin{equation}\label{eq20}
 T_{min}=(M-\textit{O})/max(M,\ceil{N/m})
\end{equation}
\normalsize
When $K=max(M,\ceil{N/m})$. In \textbf{Theorem \ref{theo2}}, when $M \geq K$, i.e., the number of available channels exceeds the number of time-slots allocated to a sensor of a \textit{WBAN}, however, the minimal throughput $T_{min}$ can be maximized when we select K equals to the maximal number of available channels, which is limited to M in our case, and so, $M < K$. Therefore, the bounds of the throughput will be impacted by the size of the Latin rectangles family \textit{m}.
\section{\textit{CHIM} Analysis}
In this section, the effectiveness of \textit{CHIM} in terms of lowering the probability of medium access collision is analyzed. 
\subsection{\textit{TDMA} Collision Probability}
We derive the probability for a designated sensor which experiences medium access collision within the \textit{TDMA} part. We assume an individual sensor $S_i$ of $\textit{WBAN}_i$ coexists in the same area with other \textit{P} sensors $S_j$, where $i \neq j$ and $S_i$ transmits a single packet in one time-slot.  $S_i$ succeeds in the transmission of its data packet to the coordinator using the default channel, \textit{if}, no sensor of the set \textit{P} transmits in the same time-slot using the same channel as $S_i$'s default channel. We represent the number of sensors that transmit their packets in the same time-slot as sensor $S_i$ as a random variable denoted by X. When x packets are transmitted in the same time-slot as $S_i$, then, the probability of event \textit{X=x} is denoted by \textit{Pr(X=x)} and defined by \textbf{eq.} \ref{eq21} below.

\small
\begin{equation}\label{eq21}
         Pr\left(X=x\right)=C_{x}^{P}\alpha^x(1-\alpha)^{P-x}\left(min(M,K)/K\right)^{x},x\leq P
\end{equation}
\normalsize
The probability for sensor $S_j$ of $\textit{WBAN}_j$ to exist within the transmission range of $\textit{WBAN}_i$ is denoted by $\alpha$. If we assume that \textit{Y} out of \textit{X} sensors decide the scheduling of their transmissions based on Latin rectangles which are orthogonal to the Latin rectangle employed by $\textit{WBAN}_i$, that means, \textit{y} out of \textit{x} sensors select symbol patterns from other orthogonal Latin rectangles that are distinct from the symbol pattern that $S_i$ uses. Thus, \textit{y} sensors will not contribute in the collision probability, and hence the probability of \textit{y} sensors that will not impose any medium access collision to $S_i$'s transmission is given by \textbf{eq.} \ref{eq22}. 

\small
\begin{equation}\label{eq22}
\begin{split}
 Pr\left(Y=y \mid X=x\right) &=\left(C_{y}^{K} C_{x-y}^{Z-K}\right)/C_{x}^{Z},\: x \leq P\: \&\: y\leq x
 \end{split}
\end{equation}
\normalsize
Where $Z = K \times m$ represents the sum of the number of symbol patterns in the orthogonal family. However, we represent the number of sensors that use the same channel as $S_i$ and collide with $S_i$'s transmission in the same time-slot as a random variable denoted by \textit{X-Y}. Thus, the probability that $S_i$'s transmission experiences a medium access collision is denoted by (\textit{collTx}) and given by \textbf{eq.} \ref{eq23} below.

\small
\begin{equation}\label{eq23}
\begin{split}
Q &= Pr(collTx\mid Y=y, X=x)\\
  &=1-Pr(succTx \mid Y=y, X=x)\\
  &=1-\left((min(M,K)-1)/min(M,K)\right)^{x-y}\\
  &=1- \left(1 - 1/min(M,K)\right)^{x-y}\\
\end{split}
\end{equation}
\normalsize
We denote all possible transmission time-slots for a sensor $S_i$ within the \textit{TDMA} part by \textit{min(M,K)} and \textit{Q} represents the probability that $S_i$ experiences a medium access collision in a single time-slot. Thus, \textit{Q} may determine the number of sensors that experience medium access collision within the \textit{TDMA} part which is denoted by \textit{W}. Then, each sensor $S_i$ $\in$ \textit{W} will use its corresponding backup channel and time-slot within the \textit{IMB} part. In the next section, we will determine the subset of backup sensors in \textit{W} that will experience interference, i.e., medium access collision within \textit{IMB} part.
\subsection{IMB Collision Probability}
We denote each sensor that occupies at least one time-slot in the \textit{IMB} part by backup sensor. In this section, we determine the probability for a backup sensor $S_i$ that experiences medium access collision when accessing its backup channel in its backup time-slot. We denote by $T_{imb}$ the total of the number of sensors that experience medium access collision along the whole \textit{TDMA} part and the number of backup sensors that experience medium access collision along the whole the \textit{IMB} part. Importantly, $T_{imb}$ follows the binomial distribution. If we assume that \textit{t} backup sensors of a designated \textit{WBAN} experience medium access collision within the \textit{IMB} part, then, the probability of the event $T_{imb} =  t$ is denoted by $Pr(T_{imb} = t)$ and is given by \textbf{eq.} \ref{eq24}.

\small
\begin{equation}\label{eq24}
        Pr(T_{imb} = t) = C_{t}^{K} (Q^{2})^t (1-Q^{2})^{K-t}, \: t\leq K 
\end{equation}
\normalsize
We add $Q^{2}$ in the equation because of the two stage medium access collision, respectively, the first occurs in the \textit{TDMA} and the second occurs in the \textit{IMB} part. Then, Q of \textbf{eq.} \ref{eq23} is substituted in \textbf{eq.} \ref{eq25}.

\small
\begin{equation}\label{eq25}
        Pr(T_{imb} = t)=C_{t}^{K} (Q^{2})^{t} (1-Q^{2})^{K-t},\: t \leq K
\end{equation}
\begin{equation}\label{eq26}
\begin{split}
 Pr(T_{imb}=t)&=C_{t}^{K}\times (Q^{2})^{t} (1-Q^{2})^{K-t} ,\: t\leq K\\
&=C_{t}^{K} \times  (1-1/min(M,K))^{(x-y)(K-t)}\\
&\times  (2-(1-1/min(M,K))^{x-y})^{K-t}\\
&\times  (1-(1-1/min(M,K)^{x-y}))^{2t}
\end{split}
\end{equation}
\normalsize
Our approach ensures that channels used in the \textit{TDMA} and the backup \textit{IMB} parts within the same superframe are always distinct. However, the superframe structure defined in the \textit{ZIGBEE} standard \cite{key5a},\cite{key5a} divides the active period into a \textit{TDMA} part and contention free period part denoted by \textit{CFP} where guaranteed time-slots denoted by \textit{GTSs} are allocated. According to this structure, a single sensor that experiences medium access collision in the \textit{TDMA} part may query its corresponding coordinator for additional \textit{GTSs} to complete its transmission. Unlike our approach, the same channel is always used in both \textit{TDMA} and \textit{CFP} parts.
\begin{lemma}
If \textit{t} sensors collide in the \textit{IMB} interference mitigation part, i.e.,  $Pr(T_{imb} = t)$, then, the probability of these sensors collide in the \textit{CFP} is given by \textbf{eq.} \ref{eq27}.
\end{lemma}

\small
\begin{equation}\label{eq27}
Pr(T_{cfp} = t) = Pr(T_{imb}=t)\times(min(M,K))^t
\end{equation}
\normalsize
\textit{\textbf{Proof:} If each \textit{WBAN} has a $M\times  K$ Latin rectangle and t sensors may face collision in the \textit{IMB} interference mitigation part, then, each sensor may have the chance to pick $min(M,K)$ possible backup channel to time-slot combinations for its transmission and hence, for $t < K$ sensors, there are $(min(M,K))^t$ possible combinations. However, in the \textit{CFP}, there is one and only one channel used by all sensors, therefore each sensor has the same channel for its transmission. Thus, in \textit{CHIM}, the probability of collisions for t sensors will be reduced by $(min(M,K))^t$ and therefore given in \textbf{eq.} \ref{eq28}}.

\begin{figure}
  \centering
   \includegraphics[width=0.3\textwidth,height=0.175\textheight]{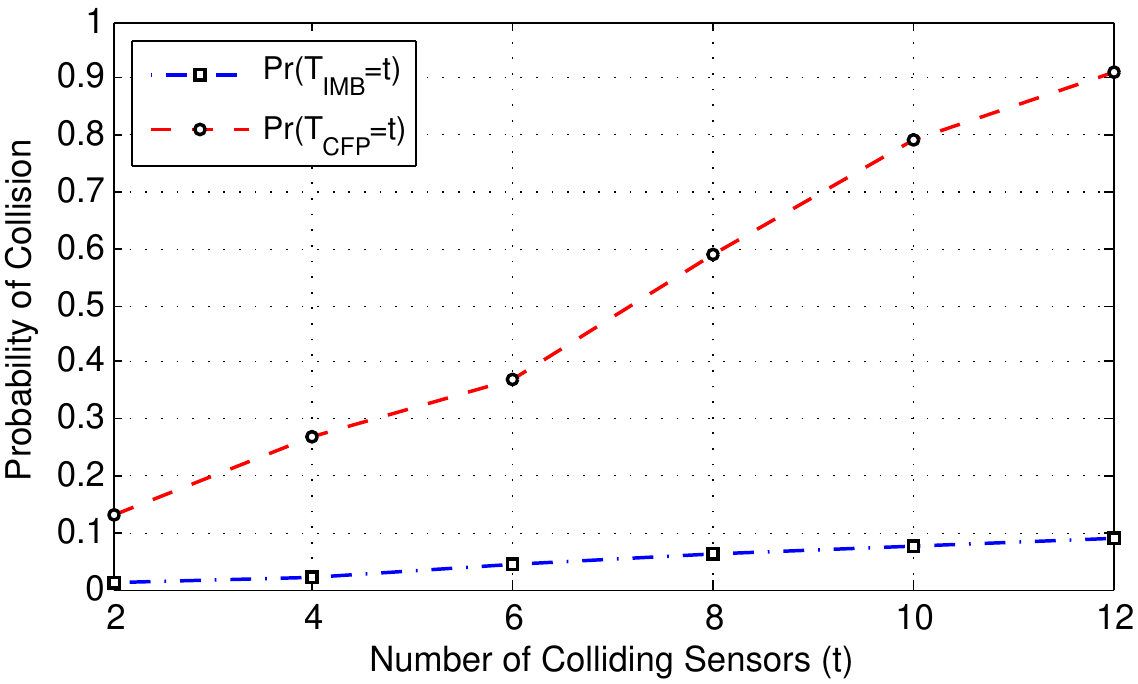}
\caption{Collision probability versus number (\#) of colliding sensors}
\label{spoc}
\end{figure}

\small
\begin{equation}\label{eq28}
Pr(T_{imb}=t) = \frac{Pr(T_{cfp} = t)}{(min(M,K))^t}
\end{equation}
\normalsize
For illustration, see \textbf{Figure \ref{spoc}}.
\section{Performance Evaluation}
We have proposed two schemes based on Latin rectangles, namely, \textit{DAIL} and \textit{CHIM}, to minimize the impact of inter-\textit{WBAN} co-channel interference through dynamic channel hopping. \textit{DAIL} assigns channel and time-slot combinations to reduce the probability of medium access collision among sensors in different \textit{WBAN}s. Yet, \textit{DAIL} involves time and power overhead due to the frequent channel hopping. Meanwhile, \textit{CHIM} takes advantage of the relatively lower density of collocated \text{WBAN}s to save power by hopping among channels only when interference is detected at the level of the individual sensors. This section compares the performance of \textit{DAIL} and \textit{CHIM} to that of competing approaches in the literature. In addition, analytical results that derive the collision probability and network throughput are validated by simulations.

We have performed simulation experiments through Matlab, where the number of \textit{WBAN}s is varied. We model a communication failure when a sensor of one \textit{WBAN} happens to pick the same channel in the same time-slot as a sensor of another \textit{WBAN}, and the two sensors are in the communication range of one another (i.e., the SNR is greater than a threshold). The locations of the individual \textit{WBAN}s change to mimic random mobility and consequently, the interference pattern varies. The following performance metrics are considered:
\begin{itemize}
\item \textbf{\textit{Collision probability (McP)}} reflects how often two transmissions of two distinct sensors of different \textit{WBAN}s collide. 
\item \textbf{\textit{Communication failure probability (CFP)}} is the frequency that two distinct sensors of different \textit{WBAN}s when both sensors are assigned the same channel in the same time. 
\item \textbf{\textit{Mean \text{WBAN} power consumption (mPC)}} is defined as the sum of the individual power consumed by the individual nodes due to the data packet collisions within a \textit{WBAN}'s superframe divided by the number of sensors in each \textit{WBAN}. 
\item \textbf{\textit{Mean successful data packets received (MsPR)}} is the total number of packets that are successfully received at the coordinator from all sensors within its \textit{WBAN} in one superframe divided by the sensor count in that \textit{WBAN}. This metric is specific for \textit{DAIL}. 
\item \textbf{\textit{Mean of deferred data packets (DPS)}} this metric is applied for \textit{CHIM} only since it provisions backup time-slots and reports the average number of transmissions that are made in backup time-slots per superframe.
\end{itemize}
\subsection{Simulation Results - \textit{DAIL}}
We have conducted multiple simulation experiments to evaluate the performance of \textit{DAIL} and compared it with that of the smart spectrum allocation scheme, denoted by \textit{SMS} \cite{key7}. \textit{SMS} assigns orthogonal channels to interfering sensors belonging to each pair of coexisting \textit{WBAN}s. The simulation parameters are provided in \textbf{Table \ref{diald}}.
\subsubsection{DAIL Collision Probability}
The \textit{first experiment} is geared for comparing the mean collision probability (\textit{McP}) versus the number of coexisting \textit{WBAN}s ($\Omega$) for \textit{DAIL} and that for \textit{SMS}. The results, shown in \textbf{Figure \ref{wbans}} confirm the advantage of \textit{DAIL} by achieving a much lower \textit{McP} because of the combined channel and time-slot hopping. It is observed that \textit{McP} of \textit{DAIL} is very low when $\Omega \leq 12$ due to the relatively large number of channel and time-slot combinations. When $12 < \Omega \leq 25$, \textit{McP} significantly grows because of the increase in the number of sensors that makes the simultaneous assignment of the same channel for more sensors is highly probable. Note that the superframe consists of up to 12 time-slots. However, when $\Omega$ exceeds 25, \textit{McP} grows very slightly and stabilizes at $21 \times 10^{-2}$ eventually as the number of medium access collisions reaches its maximum bound by each \textit{WBAN} due to the limited size of the Latin rectangle in terms of channels and time-slot combinations. In other words, all \textit{M} channels in the Latin rectangle are assigned sensors and hence, many in other \textit{WBAN}s are competing for these channels in the same time. Thus, the number of competing sensors for all channels is significantly larger than the number of available channels and so the collision reaches its maximum level. Meanwhile, in \textit{SMS}, \textit{McP} significantly increases when $0 < \Omega \leq 18$, i.e., the number of available orthogonal channels is smaller than the number of interfering sensors. Then, \textit{McP} slightly increases until it stabilizes at $5 \times 10^{-1}$ when $18 < \Omega \leq 35$ since the interference attains its maximum and all channels are already assigned, i.e., the number of collisions attained by each \textit{WBAN} is larger than the available channels. \textit{McP} significantly grows for as long as the number of channels is smaller than $\Omega$. However, when $\Omega$ exceeds \textit{16}, \textit{McP} tends to stabilize at $5 \times 10^{-1}$.
\begin{table}
\centering
\caption{Simulation parameters - \textit{DAIL}}
\label{diald}\resizebox{0.45\textwidth}{!}{
\begin{tabular}{llllll}
\noalign{\smallskip}\hline
\hline\noalign{\smallskip}
&\textbf{Exp. 1} & \textbf{Exp. 2} &\textbf{Exp. 3}&\textbf{Exp. 4}\\
Sensor TxPower(dBm) & -10 &-10 &-10&-10\\
\# Coordinators/\textit{WBAN} & 1 &1 &1 & 1 \\
\# Sensors/\textit{WBAN} &12 &12 & 12&12\\
\# \textit{WBAN}s/Network &Var & 30 & Var&Var\\
\# Time-slots/Superframe  &12 &12 &12& 12\\
Latin Rectangle Size&$16 \times 12$ &$16 \times Var$ &$16 \times 12$& $16 \times 12$\\
\noalign{\smallskip}\hline
\hline\noalign{\smallskip}
\end{tabular}}
\end{table}
Our \textit{second experiment} studies the effect of the number of time-slots per a superframe denoted by \textit{TL} on \textit{McP} for a network consisting of up to 30 coexisting \textit{WBAN}s. As can be clearly seen in \textbf{Figure \ref{slots}}, \textit{DAIL} always achieves lower collision probability than \textit{SMS} for all \textit{TL} values. In \textit{DAIL}, \textit{McP} significantly decreases as \textit{TL} increases from $10$ to $28$, where increasing \textit{TL} is similar to enlarging the size of the Latin rectangle. Therefore, a larger number of channel and time-slot combinations allows distinct sensors to not pick the same channel in the same time-slot, which decreases the chances of collisions among them. However, \textit{SMS} depends only on the \textit{16} available channels to mitigate interference, and the channel assigned to a sensor stays the same at all time. Thus, a high \textit{McP} is expected due to the larger number of interfering sensors than the number of available channels. Moreover, a sensor has \textit{16} choices in \textit{SMS}, while it has $16 \times framesize$ different choices in \textit{DAIL} to mitigate the interference, which explains the large difference in \textit{McP} amongst the two schemes.
\subsubsection{DAIL Power Consumption}
In \textit{third experiment}, we compare the mean power consumption denoted by \textit{mPC} of each \textit{WBAN} versus the number of coexisting \textit{WBAN}s ($\Omega$) for \textit{DAIL} and \textit{SMS}. The results plotted in \textbf{Figure \ref{ec}} show that \textit{mPC} for \textit{DAIL} is always lower than that of \textit{SMS} for all $\Omega$ values. \textit{DAIL} exposes such performance because of the reduction in the number of medium access collisions which lead to a fewer number of retransmissions and hence high energy savings. For \textit{DAIL}, it is shown in the figure that \textit{mPC} slightly grows when $\Omega \leq 10$ because the number of channel and time-slot combination is larger than the number of the interfering sensors, which reduces the number of medium access collisions amongst the coexisting sensors. \textit{mPC} significantly increases when $10 < \Omega \leq 30$ because of the larger number of sensors simultaneously compete to the same channel which results in a larger number of medium access collisions and hence the power consumption is increased accordingly. When $\Omega$ exceeds 30, the power consumption increases slightly to stabilize at $16.5 \times 10^{-3}mW$ because of the medium contention reaches its maximum level and no additional power is introduced due to the maximal number of medium access collisions reached by each \textit{WBAN}. Meanwhile, in \textit{SMS}, \textit{mPC} is consistently high for large networks due to the medium access collisions resulting from the large number of sensors that compete for the available channels (\textit{16 channels}).
\subsubsection{DAIL WBAN Throughput}
\begin{figure*}
\begin{minipage}[b]{.305\textwidth}
\centering
\includegraphics[width=1\textwidth, height=0.2\textheight]{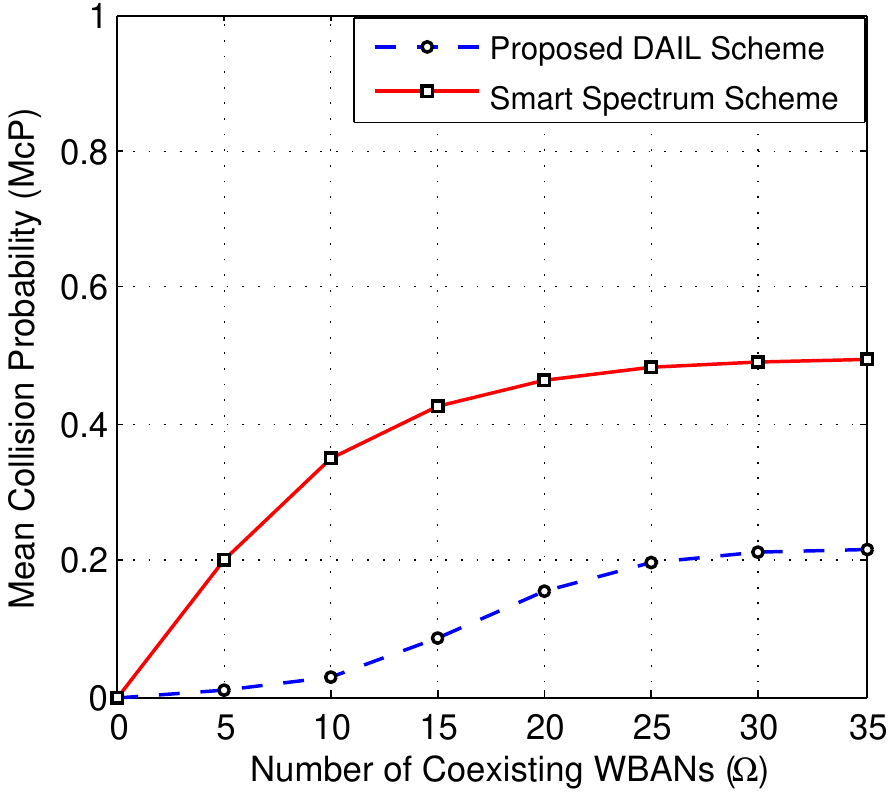}
\caption{Mean collision probability (\textit{McP}) versus $\Omega$}
\label{wbans}
\end{minipage}\qquad
\begin{minipage}[b]{.305\textwidth}
\centering
\includegraphics[width=1\textwidth, height=0.2\textheight]{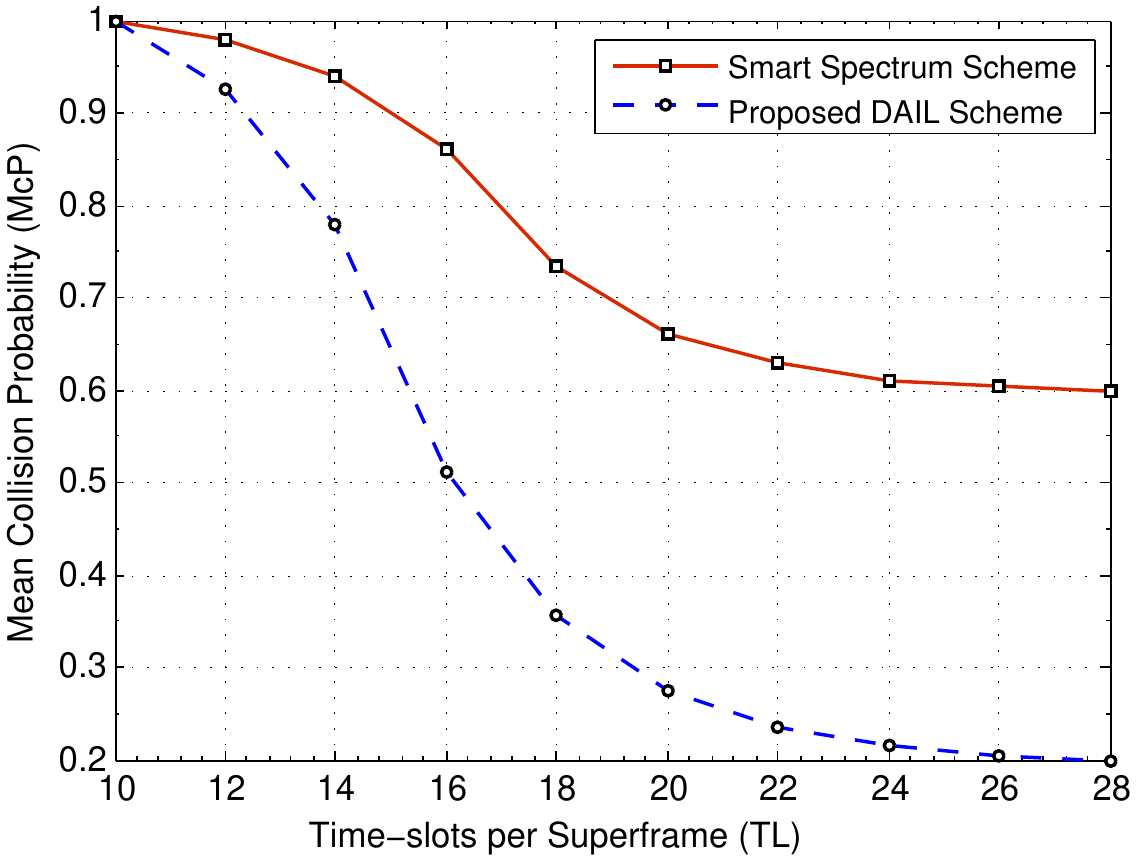}
\caption{Mean collision probability (\textit{McP}) versus \# of time-slots per superframe}
\label{slots}
\end{minipage}\qquad
\begin{minipage}[b]{.305\textwidth}
\centering
\includegraphics[width=1\textwidth, height=0.2\textheight]{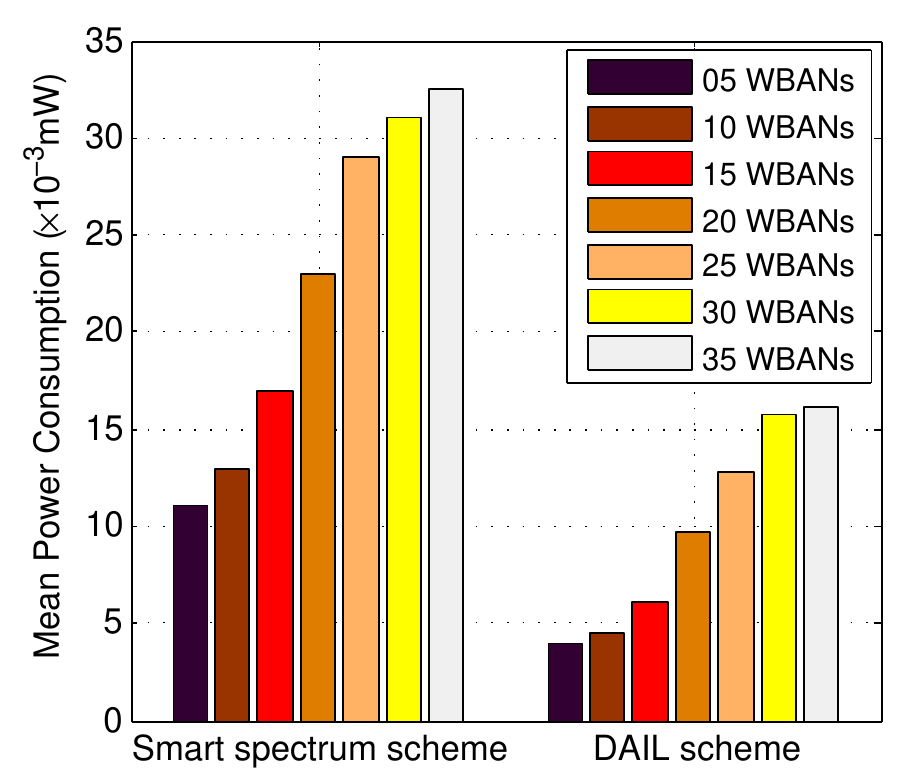}
\caption{Mean power consumption (\textit{mPC}) versus $\Omega$}
\label{ec}
\end{minipage}\qquad
\end{figure*}

\begin{figure*}
\begin{minipage}[b]{.305\textwidth}
\centering
\includegraphics[width=1\textwidth, height=0.2\textheight]{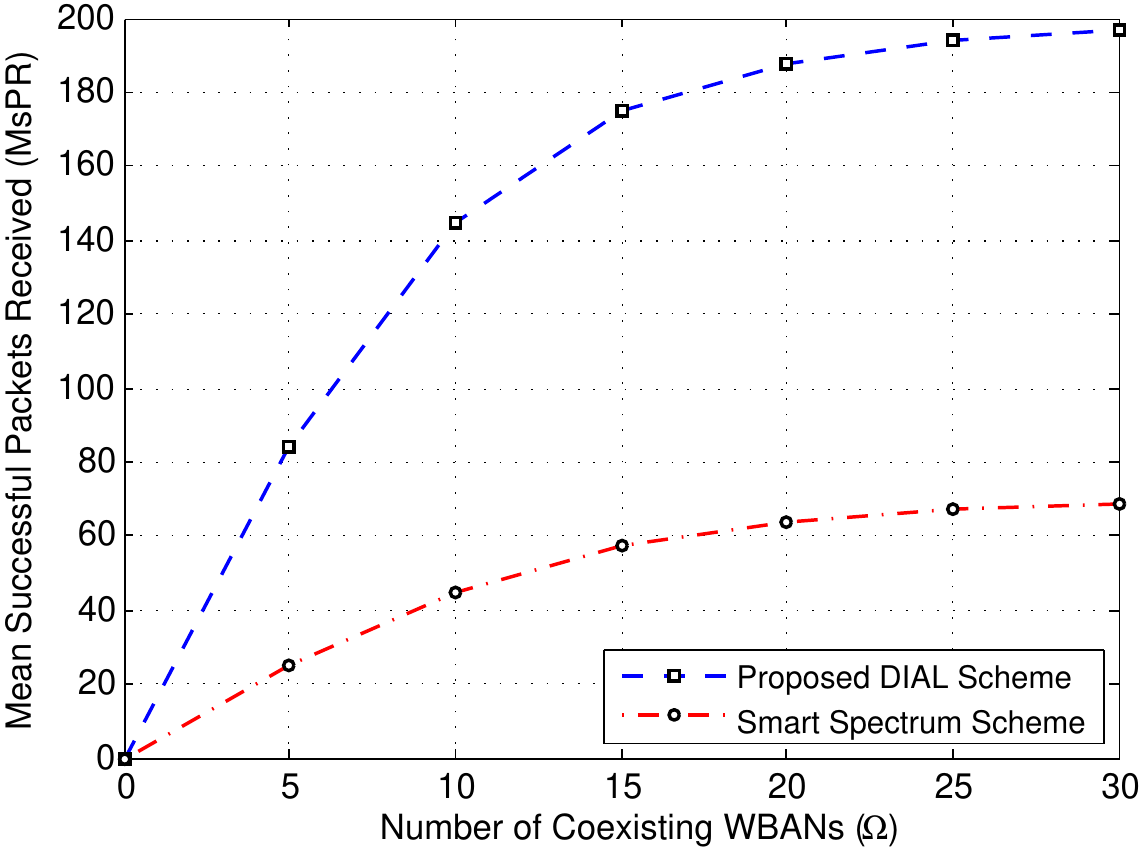}
\caption{Mean successful packets received (\textit{MsPR}) versus $\Omega$}
\label{thro}
\end{minipage}\qquad
\begin{minipage}[b]{.305\textwidth}
\centering
\includegraphics[width=1\textwidth, height=0.2\textheight]{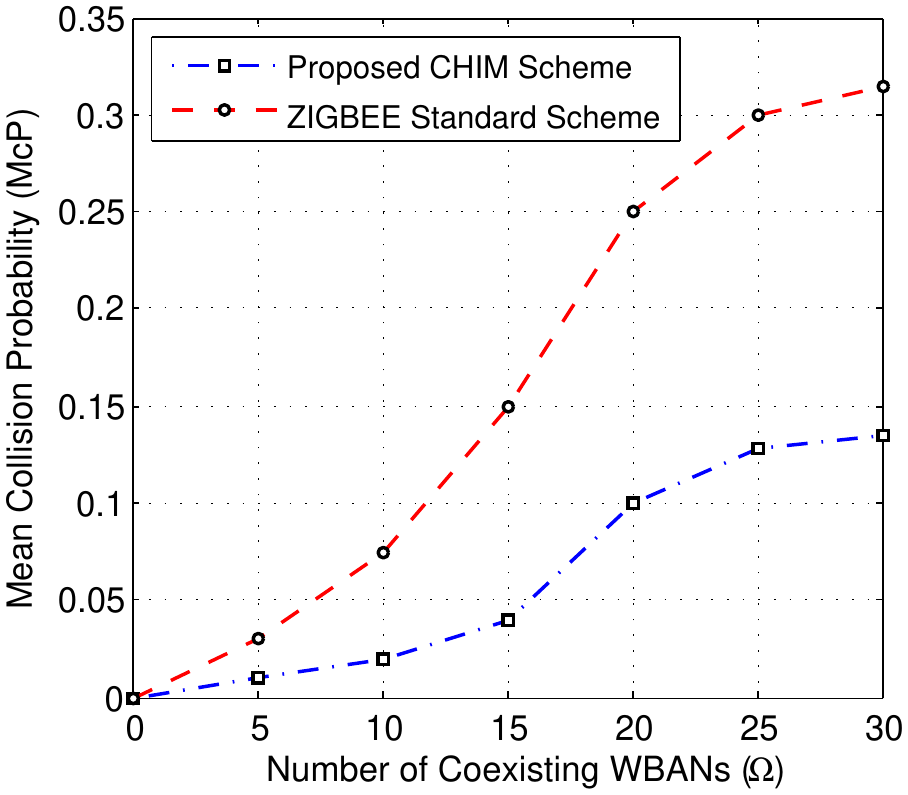}
\caption{Mean collision probability (\textit{McP}) versus $\Omega$}
\label{FOC}
\end{minipage}\qquad
\begin{minipage}[b]{.305\textwidth}
\centering
\includegraphics[width=1\textwidth, height=0.2\textheight]{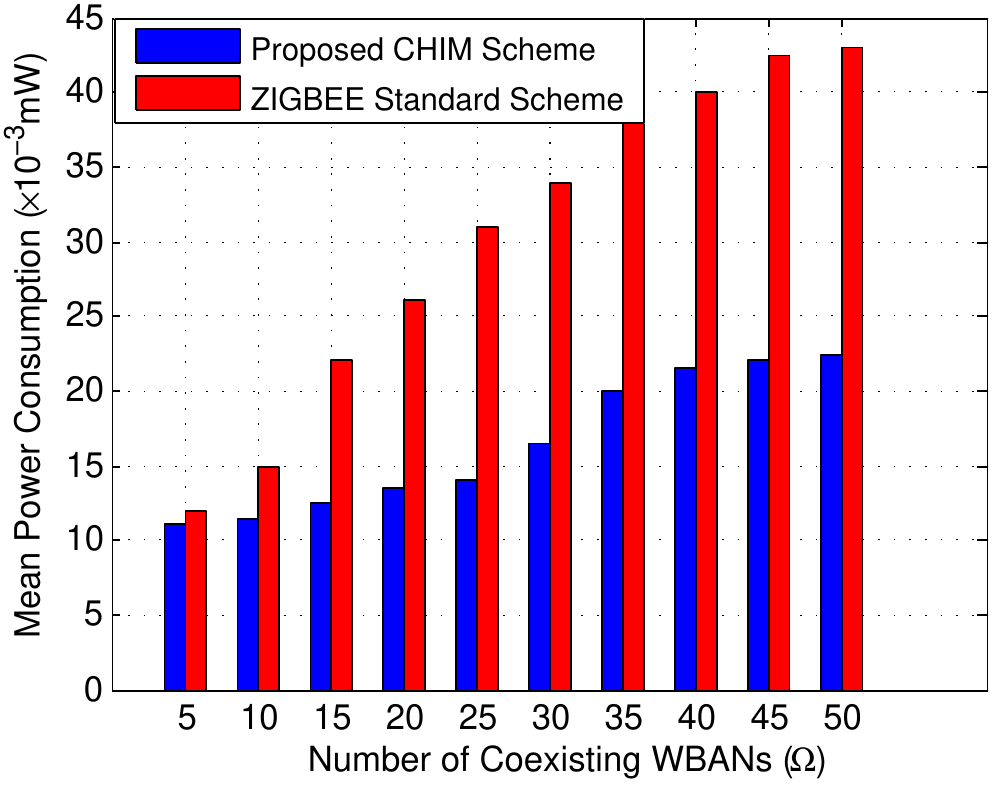}
\caption{Mean power consumption (\textit{mPC}) versus $\Omega$}
\label{AEC}
\end{minipage}\qquad
\end{figure*}

The \textit{fourth experiment} studies the mean successful data packets received at each \textit{WBAN}, denoted by \textit{MsPR}, versus $\Omega$ for \textit{DAIL} and \textit{SMS}. \textbf{Figure \ref{thro}} shows that \textit{DAIL} always achieves higher \textit{MsPR} than \textit{SMS} for all values of $\Omega$. Such performance improvement is mainly because of the reduced number of medium access collisions, which boosts the number of data packets that are successfully received in a superframe. For \textit{DAIL}, the figure shows that \textit{MsPR} significantly increases when $\Omega \leq 10$ due to the lower number of collisions; as we explained when discussing \textbf{Figure \ref{colors}}. \textit{MsPR} slightly increases when $\Omega > 10$ due to the large number of sensors competing for the same channel in the same time-slots, which leads to a larger number of medium access collisions. When $\Omega$ exceeds 25, \textit{MsPR} stabilizes at $198$ because of the medium contention reaches its maximum level as shown in \textbf{Figure \ref{wbans}}. The throughput for \textit{SMS} is also consistent with the \textit{McP} results in \textbf{Figure \ref{wbans}}. Basically, in \textit{SMS}, \textit{MsPR} significantly increases as $\Omega$ grows for as long as $\Omega \leq 15$ due to the availability of a larger the number of channels than the number of interfering sensors. For $15 < \Omega \leq 25$, a low \textit{MsPR} is observed since many sensors compete for available \textit{16} channels. When $\Omega$ exceeds 25, \textit{MsPR} increases very slightly and eventually stabilizes at 67, because none of the channels is available to be assigned for an interfering sensor.
\begin{table}
\centering
\caption{Simulation parameters - \textit{CHIM}}
\label{parm}
\resizebox{0.3\textwidth}{!}{
\begin{tabular}{lll}
\noalign{\smallskip}\hline
\hline\noalign{\smallskip}
 Sensor\textit{TxPower(dBm)} & -10\\
 Sensors/\textit{WBAN} &20\\
 \textit{WBAN}s/Network &Variable \\
 Time-slots/\textit{TDMA} \textit{CHIM} part  &20\\
 Time-slots/\textit{IMB \textit{CHIM}} part  &20\\
 Time-slots/\textit{TDMA} \textit{ZIGBEE} part  &20\\
 Time-slots/\textit{CFP} \textit{ZIGBEE} part  &20\\
 Latin Rectangle Size&$16 \times 20$\\
\noalign{\smallskip}\hline
\hline\noalign{\smallskip}
\end{tabular}}
\end{table}
\subsection{Simulation Results - \textit{CHIM}}
We have conducted multiple experimental simulations to study the performance of \textit{CHIM} using the same simulation environment. Unlike \textit{DAIL}, the performance of \textit{CHIM} is compared with \textit{ZIGBEE} standard \cite{key5a} since it resembles \textit{CHIM} in terms of using one channel for intra-\textit{WBAN} communication. The \textit{ZIGBEE} standard assigns \textit{GTSs} in (\textit{CFP}) to the sensors experienced co-channel interference in the \textit{TDMA} period. The experimental parameters are provided in \textbf{Table \ref{parm}}.
\subsubsection{CHIM Collision Probability}
The effect of the variable $\Omega$ on \textit{McP} for \textit{CHIM} and \textit{ZIGBEE} is reported in \textbf{Figure \ref{FOC}}. It is clearly shown in the figure, \textit{CHIM} provides a much lower \textit{McP} is mainly because of the channel hopping. As can be seen on this figure, \textit{McP} for \textit{CHIM} is very low when $\Omega \leq 15$, because of the large number of the channel hopping possibilities in the \textit{IMB} frame. When $15 < \Omega \leq 25$, \textit{McP} significantly increases because of the larger number of sensors which makes the possibility of more sensors to be assigned simultaneously the same channel is highly probable. When $\Omega$ exceeds 25, \textit{McP} grows very slightly and eventually stabilizes at $135 \times 10^{-3}$ because of the limited availability of the Latin rectangles and hence, the limited availability of the backup channels/time-slots. In other words, all the \textit{TDMA} and backup time-slots/channels are completely committed, and any competing sensor that have data to transmit will face collision and hence the medium capacity reaches its maximum level. In \textit{ZIGBEE}, \textit{McP} slightly grows when $0 < \Omega \leq 10$ because of the number of available \textit{GTS} time-slots is similar to the number of interfering sensors. Then, when $10 < \Omega \leq 25$, \textit{McP} significantly grows due to the increase in the number of interfering sensors. \textit{McP} stabilizes at $315 \times 10^{-2}$ when $\Omega \geq 25$ because of the large number of the interfering sensors than the number of the available \textit{GTS} time-slots.
\subsubsection{CHIM Power Consumption}
\textbf{Figure \ref{AEC}} shows the mean power consumption (\textit{mPC}) of a \textit{WBAN} versus $\Omega$ for \textit{CHIM} and \textit{ZIGBEE}. As can be observed from \textbf{Figure \ref{AEC}}, \textit{CHIM} always achieves a lower \textit{mPC} than that of \textit{ZIGBEE} for all values of $\Omega$. \textit{CHIM} outperforms \textit{ZIGBEE} because of the reduced number of medium access collisions which results in a smaller number of packet retransmissions and consequently lower \textit{mPC}. For \textit{CHIM}, the figure shows a trend which is consistent with \textbf{Figure \ref{FOC}}. Basically, \textit{mPC} slightly grows when $\Omega \leq 20$ because there exists a large number of hopping choices exceeding the number of interfering sensors which decreases the number of medium access collisions amongst them and in consequence \textit{mPC}. When $\Omega$ exceeds 40, \textit{mPC} grows slightly to stabilize at $22.5 \times 10^{-3}mW$ because of the limited number of the available backup channels and time-slot combinations. However, in \textit{ZIGBEE}, \textit{mPC} slightly increases when $0 < \Omega \leq 10$ because of the similar number of interfering sensors and available \textit{GTSs}, which makes it lowly probable for more sensors to use the same \textit{GTS}. When $10 < \Omega \leq 40$, \textit{mPC} significantly grows because of the increase in the number of interfering sensors. \textit{mPC} stabilizes at $43 \times 10^{-3}mW$ when $\Omega \geq 45$ because of the lower number of available \textit{GTS} time-slots than the number of the interfering sensors.
\subsubsection{CHIM Throughput}
\textbf{Figure \ref{delay}} compares the mean number of deferred data packets (\textit{DPS}) for \textit{CHIM} and \textit{ZIGBEE} based on \textit{20} coexisting \textit{WBAN}s and while varying the number of transmitted superframes. The figure shows the values of \textit{DPS} for \textit{CHIM} is always smaller than that of \textit{ZIGBEE} which is mainly because of the reduced number of sensors contending to the medium which results in a lower of number of deferred data packets and consequently lower transmission delay. Accordingly, the throughput is increased. As \textit{ZIGBEE} uses one channel rather than \textit{16}, \textit{ZIGBEE} provides higher \textit{DPS} than that of \textit{CHIM} and hence the number of sensors which compete to the same \textit{GTS} is large, that results in a higher number of deferred packets and consequently the throughput is degraded.
\subsection{Comparing \textit{DAIL} and \textit{CHIM}}
We also conducted simulation experiments to compare the performance of \textit{DAIL} and \textit{CHIM}. We have studied the effect of the number of \textit{WBAN}s on collision and communication failure probabilities of sensor transmission, \textit{WBAN} power consumption and throughput. The simulation parameters for both \textit{DAIL} and \textit{CHIM} are provided in \textbf{Table \ref{dailchim}}.
\begin{table}
\centering
\caption{Simulation parameters - \textit{DAIL \& CHIM}}
\label{dailchim}
\resizebox{0.3\textwidth}{!}{
\begin{tabular}{lll}
\noalign{\smallskip}\hline
\hline\noalign{\smallskip}
&\textbf{DAIL} & \textbf{CHIM}\\
Sensor TxPower(dBm) & -10 &-10 \\
\# Coordinators/\textit{WBAN} & 1 &1  \\
\# Sensors/\textit{WBAN} &20 &20 \\
\# \textit{WBAN}s/Network &Var & Var\\
\# Time-slots/Superframe  &40 &40 \\
Latin Rectangle Size&$16 \times 20$ &$16 \times 20$ \\
\noalign{\smallskip}\hline
\hline\noalign{\smallskip}
\end{tabular}}
\end{table}
\subsubsection{Collision Probability}
\begin{figure*}
\begin{minipage}[b]{.305\textwidth}
\centering
\includegraphics[width=1\textwidth, height=0.2\textheight]{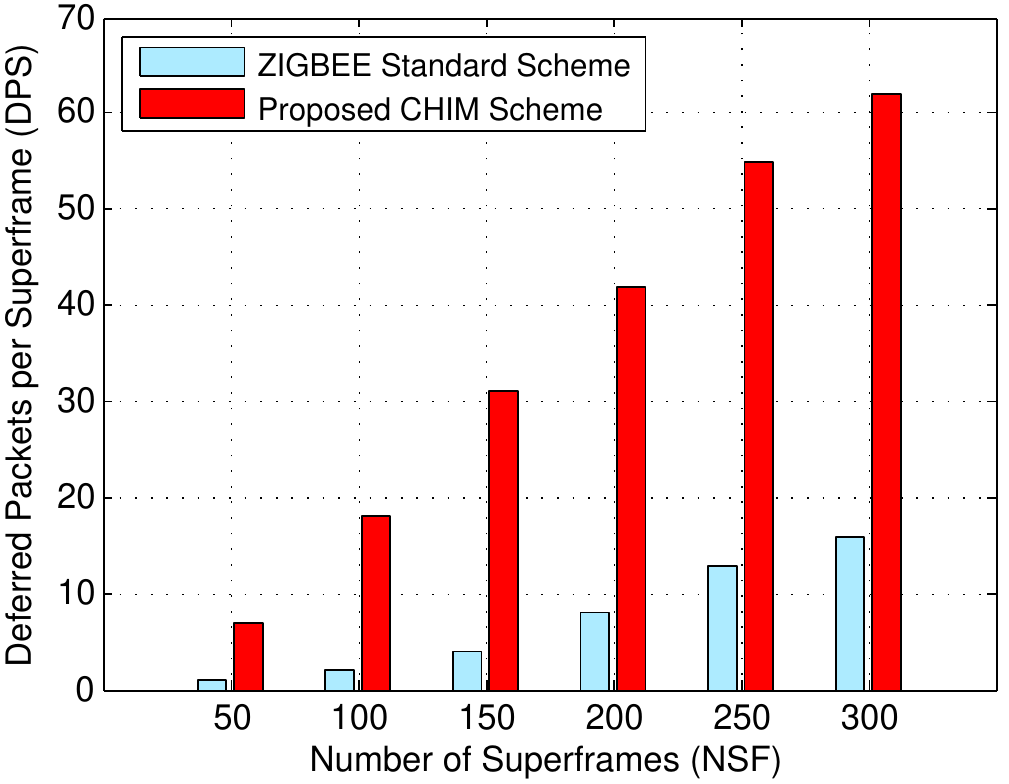}
\caption{Mean \# of packets (\textit{DPS}) versus \# of transmitted superframes}
\label{delay}
\end{minipage}\qquad
\begin{minipage}[b]{.305\textwidth}
\centering
\includegraphics[width=1\textwidth,height=0.2\textheight]{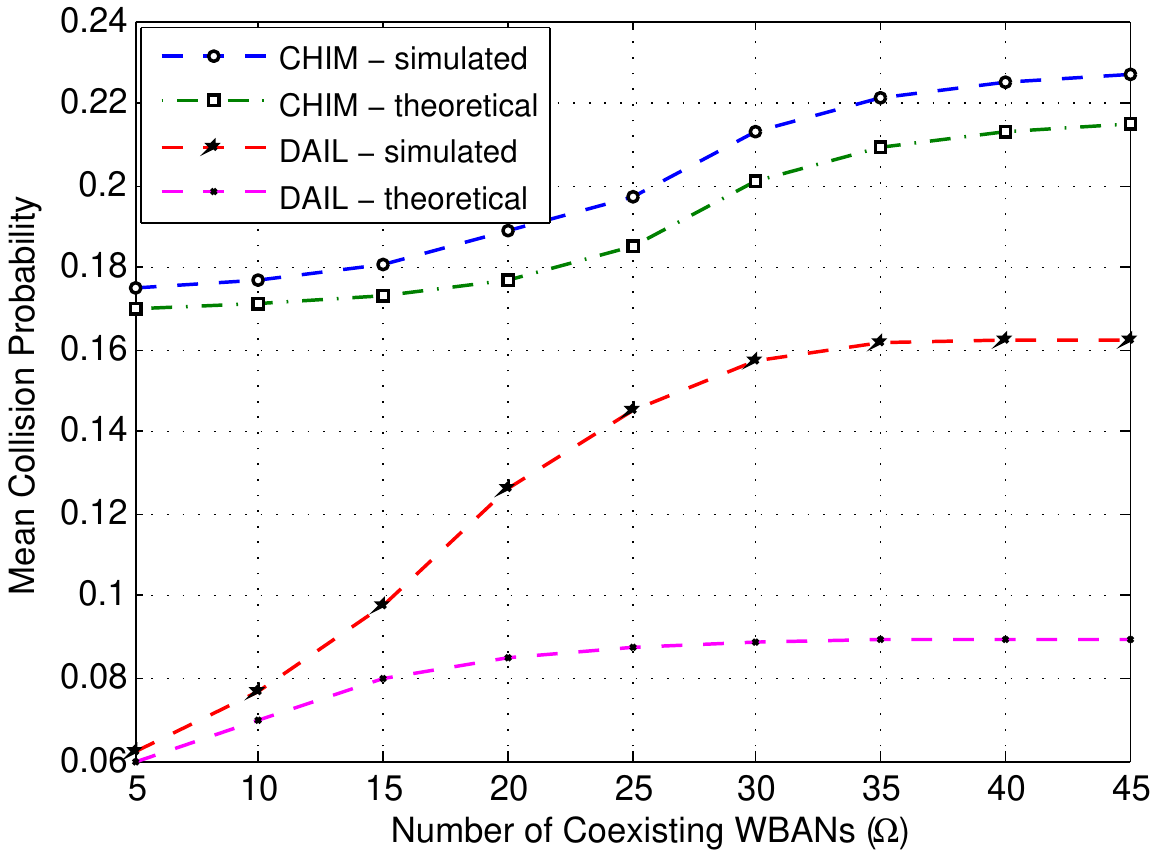}
\caption{Mean collision probability (McP) versus \# of \text{WBAN}s ($\Omega$)}
\label{cpc}
\end{minipage}\qquad
\begin{minipage}[b]{.305\textwidth}
\centering
\includegraphics[width=1\textwidth, height=0.2\textheight]{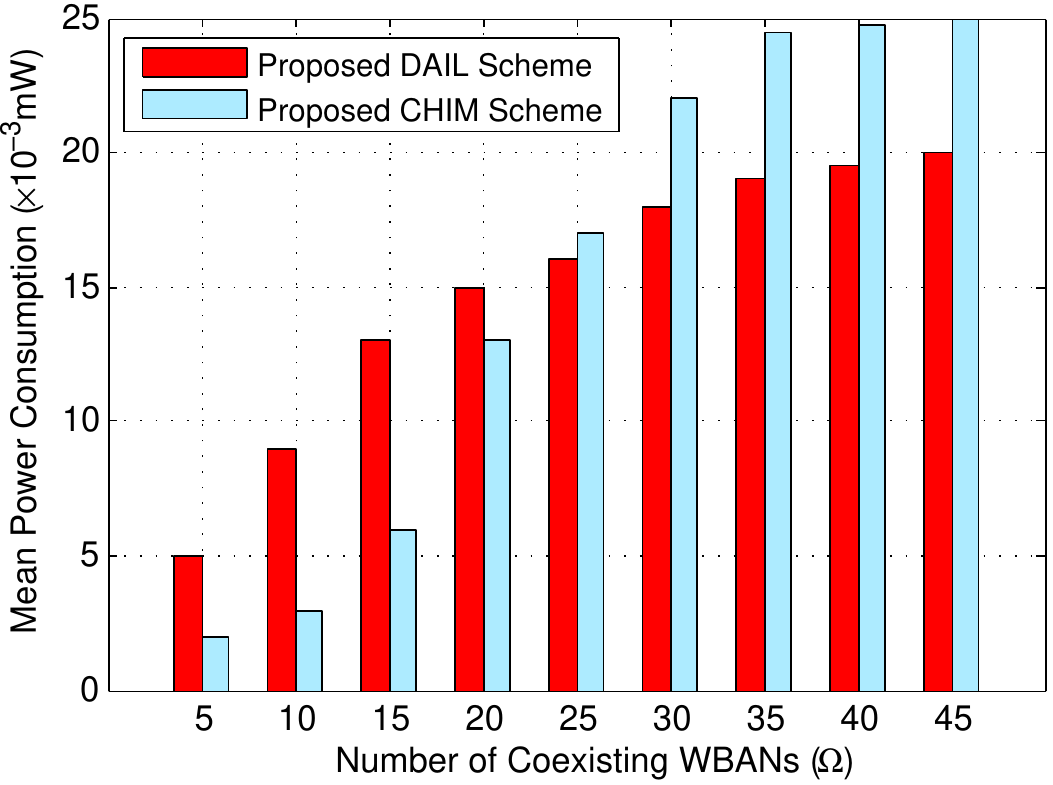}
\caption{Mean power consumption (\textit{mPC}) versus $\Omega$ for \textit{DAIL} and \textit{CHIM}}
\label{ecd}
\end{minipage}\qquad
\end{figure*}

The theoretical and simulated mean collision probability (\textit{McP}) versus $\Omega$ for \textit{DAIL} and \textit{CHIM} are compared in \textbf{Figure \ref{cpc}}. As seen in the figure, for both \textit{DAIL} and \textit{CHIM}, the simulated \textit{McP} is always higher than the theoretical \textit{McP} for all values of $\Omega$. This is due to the larger number of coexisting \textit{WBAN}s (which is made dynamic) in the simulation setup than in the theoretical study (which is made a constant). The mean collision probability (\textit{McP}) versus $\Omega$ for \textit{DAIL} and \textit{CHIM} are compared in \textbf{Figure \ref{cpc}}. As can be seen on the figure, \textit{DAIL} always provides a lower \textit{McP} than that of \textit{CHIM} for all values of $\Omega$. It is observed from this figure that \textit{McP} of \textit{DAIL} significantly increases when $\Omega \leq 25$ because of the increase in the number of sensors that makes the possibility of more sensors to be assigned simultaneously the same channel is highly probable and it slightly increases when $\Omega$ exceeds 25 until it eventually stabilizes at 0.162 due to the limited availability of orthogonal Latin rectangles. However, \textit{McP} of \textit{CHIM} is low when $\Omega \leq 15$ due to the availability of sufficient number of distinct channels. \textit{McP} significantly increases when $\Omega$ exceeds 15 until it eventually stabilizes at 0.228 due to the single channel used in the TDMA frame and the limited availability of orthogonal Latin rectangles and hence, the number of collisions is larger than the number of available backup time-slots/channels. Therefore, from a design point of view, \textit{DAIL} is consistently better than \textit{CHIM} in terms of collision probability.
\begin{figure*}
\begin{minipage}[b]{.305\textwidth}
\centering
\includegraphics[width=1\textwidth,height=0.2\textheight]{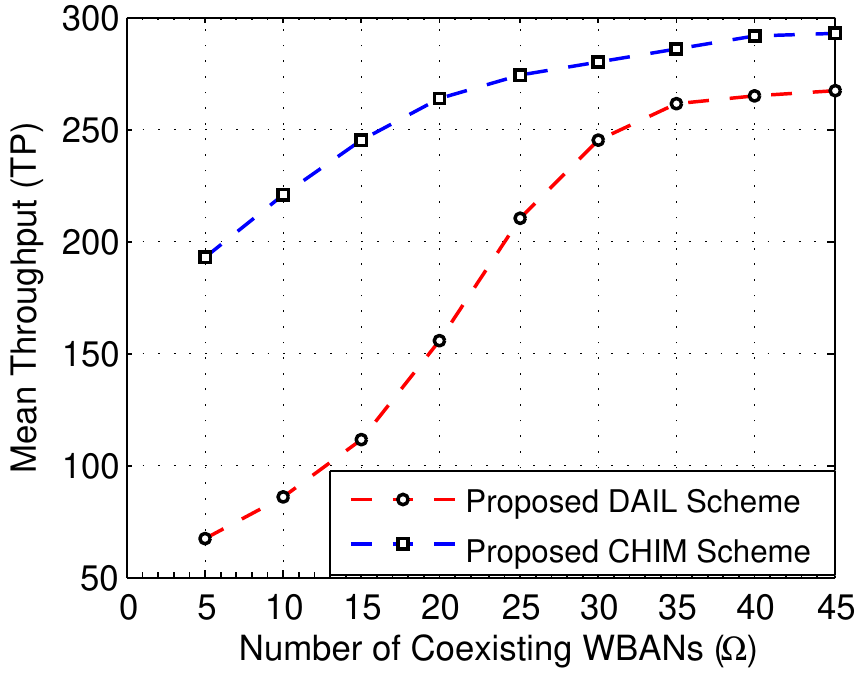}
\caption{Mean throughput (\textit{TP}) versus \# of \text{WBAN}s ($\Omega$)}
\label{thrpt}
\end{minipage}\qquad
\begin{minipage}[b]{.305\textwidth}
\centering
\includegraphics[width=1\textwidth,height=0.2\textheight]{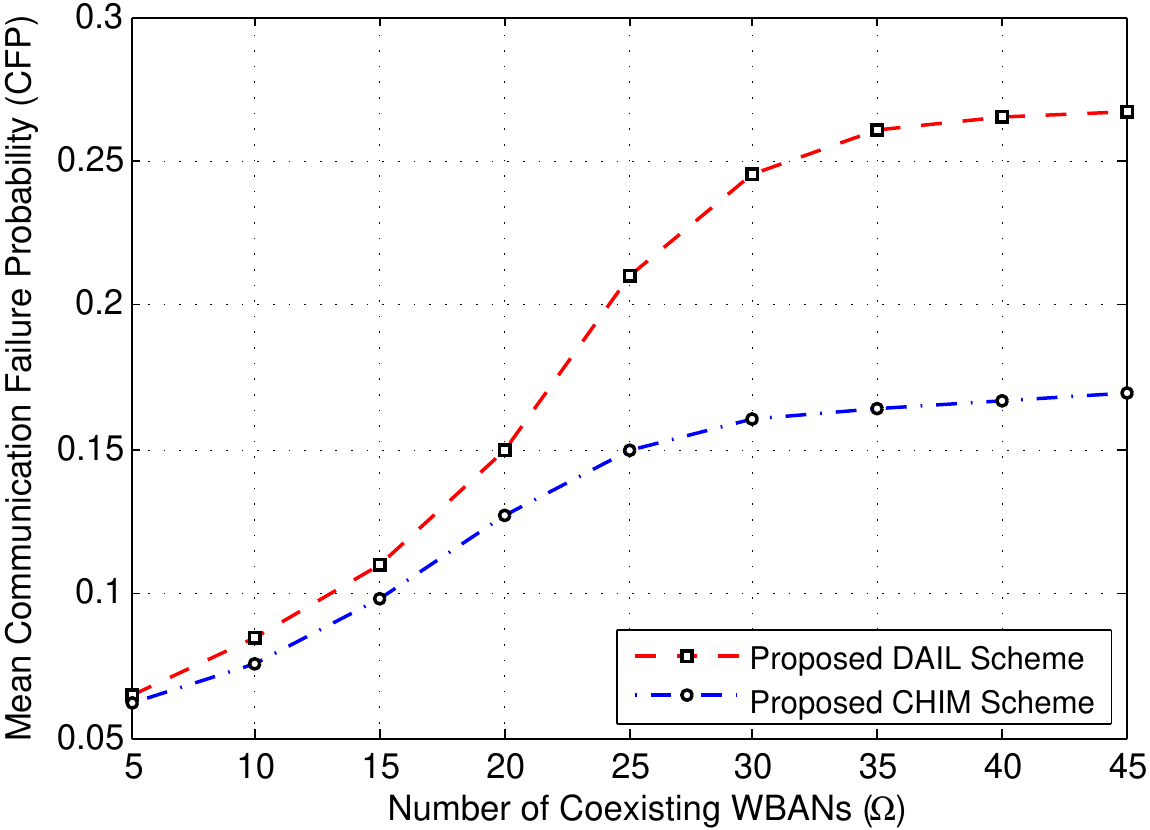}
\caption{Mean communication failure probability (CFP) versus \# of \text{WBAN}s ($\Omega$)}
\label{cfpdc}
\end{minipage}\qquad
\begin{minipage}[b]{.305\textwidth}
  \centering
   \includegraphics[width=1\textwidth,height=0.2\textheight]{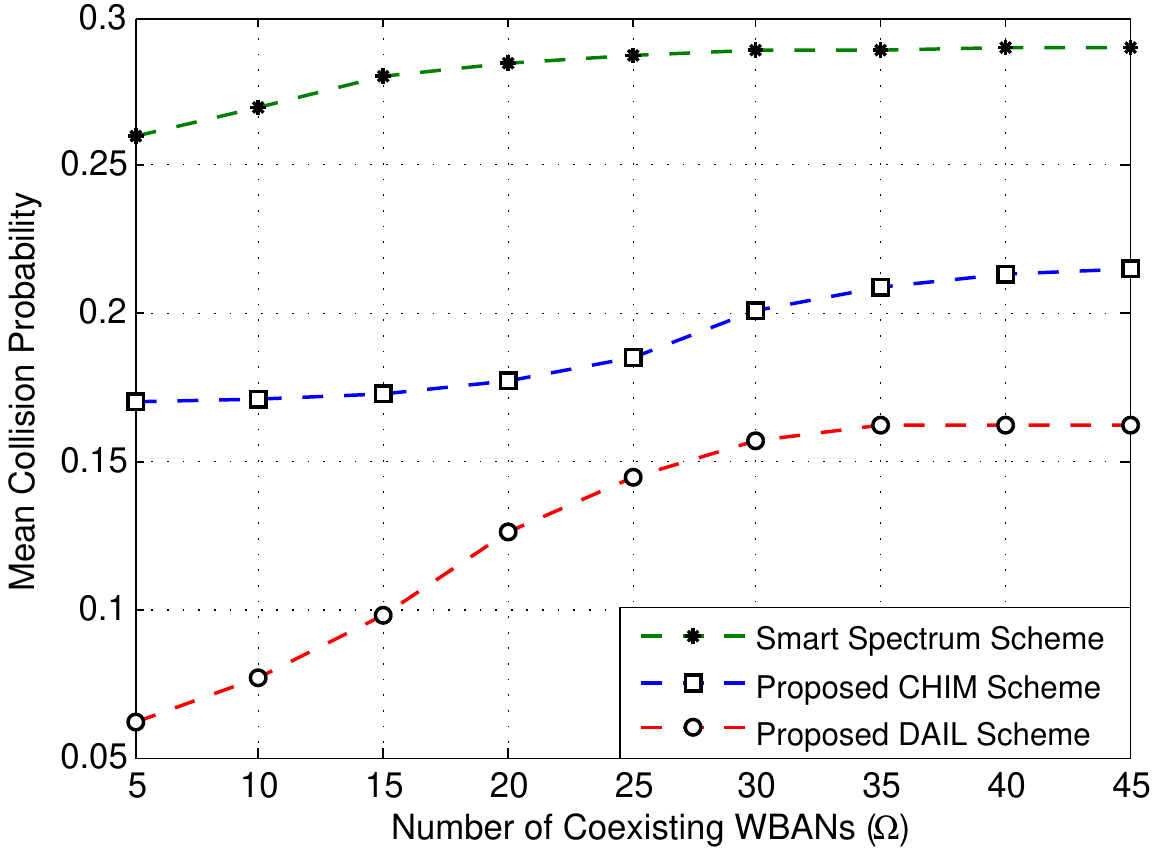}
\caption{Collision probability versus \# of coexisting \textit{WBAN}s}
\label{scs}
\end{minipage}\qquad
\end{figure*}
\subsubsection{Power Consumption}
The mean power consumption (\textit{mPC}) versus $\Omega$ for \textit{DAIL} and \textit{CHIM} are compared. \textbf{Figure \ref{ecd}} shows that \textit{mPC} for \textit{DAIL} is larger than for \textit{CHIM} when $\Omega \leq 20$ because all sensors in \textit{DAIL} need to hop among the channels, each within its assigned time-slot regardless of there is interference or not, while, in \textit{CHIM}, a sensor only switches the channel when it experiences an interference. 
\begin{figure}
  \centering
   \includegraphics[width=0.3\textwidth, height=0.175\textheight]{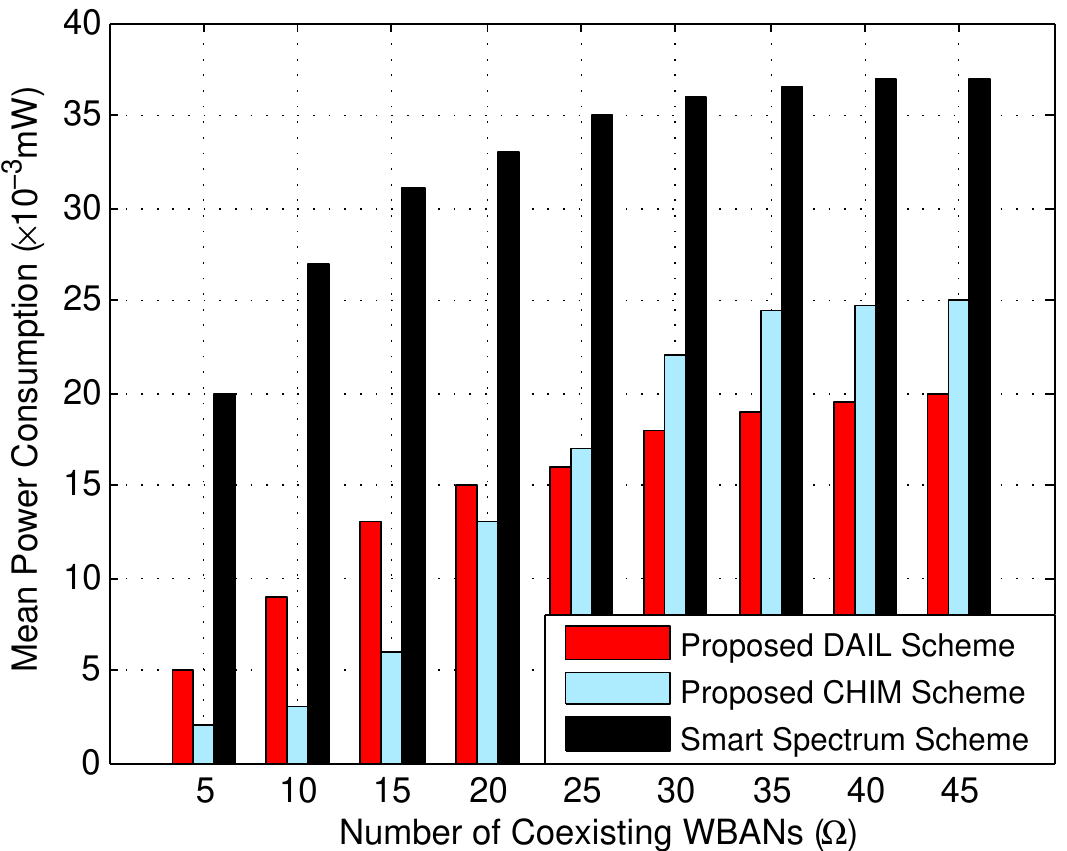}
\caption{Mean power consumption versus \# of coexisting \textit{WBAN}s}
\label{energydcs}
\end{figure}
When $\Omega$ exceeds 20, \textit{mPC} for \textit{CHIM} is higher than that of \textit{DAIL} because of the increased number of collisions (and consequently retransmissions) experienced due to the limited availability of the channels. In \textit{DAIL}, the power consumption is accumulated from the power consumed due to the high frequency of channel switching that results from the frequent channel hopping \cite{key5}. Therefore, from power consumption point of view, \textit{CHIM} suits relatively low density of \textit{WBAN}s, whilst, \textit{DAIL} suits crowded area of \textit{WBAN}s.
\subsubsection{Throughput}
The throughput (\textit{TP}) for \textit{DAIL} and \textit{CHIM} is reported in \textbf{Figure \ref{thrpt}} as a function of the number of coexisting \text{WBAN}s ($\Omega$). \textbf{Figure \ref{thrpt}} shows that \textit{CHIM} always achieves higher \textit{TP} than \textit{DAIL} for all values of $\Omega$. Such high throughput in \textit{CHIM} is mainly because of the reduced collisions and availability of backup time-slots, which boosts the number of data packets that are successfully received in a superframe. When $\Omega$ exceeds 20, \textit{TP} of \textit{CHIM} eventually stabilizes at 298 due to the high communication failure probability and the limited availability of orthogonal Latin rectangles, which degrade the effectiveness of the backup time-slots/channels. However, \textit{DAIL} always achieves lower \textit{TP} than \textit{CHIM} for all values of $\Omega$ due to the absence of backup time-slots/channels and the limited availability of orthogonal Latin rectangles which make the probability of multiple sensors pick the same channel in the same time is low.
\subsubsection{Communication Failure Probability}
The mean communication failure probability \textit{CFP} versus the number of coexisting \textit{WBAN}s ($\Omega$) for \textit{DAIL} and \textit{CHIM} are compared in \textbf{Figure \ref{cfpdc}}. As can be seen on the figure, \textit{CHIM} and \textit{DAIL} yield similar low \textit{CFP} when $\Omega \leq 15$ due to the availability of channel choices and backup time-slots/channels more than the number of interfering sensors. When $\Omega$ exceeds 15, the \textit{CFP} of \textit{DAIL} grows significantly until it eventually stabilizes at 0.227 due to the limited availability of orthogonal Latin rectangles which makes the probability of two or more sensor collide in the same time-slot is high, i.e., when the number of interfering sensors is significantly larger than the number of channel and time-slot combinations. However, when $\Omega$ exceeds 15, \textit{CFP} of \textit{CHIM} significantly increases until it eventually stabilizes at 0.17 due to the high communication failure probability and the limited availability of orthogonal Latin rectangles.
\subsection{Comparing \textit{DAIL}, \textit{CHIM} and \textit{SMS}}
\subsubsection{Collision Probability}
The mean collision probability (\textit{McP}) versus $\Omega$ for \textit{DAIL}, \textit{CHIM} and smart spectrum scheme (\textit{SMS}) are compared in \textbf{Figure \ref{scs}}. As seen in the figure, for both \textit{DAIL} and \textit{CHIM}, the \textit{McP} is always lower than \textit{SMS} for all values of $\Omega$ because of the large number of channel and time-slot combinations and the backup time-slots that prevent distinct sensors to pick the same channel at the same time-slot. This large number of combinations reduces the chances of medium access collisions among the coexisitng \textit{WBAN}s. However, \textit{SMS} depends only on the \textit{16} available channels to mitigate interference, and the channel assigned to a sensor stays the same at all time. Thus, a high \textit{McP} is expected due to the larger number of interfering sensors than the number of available channels. Moreover, a sensor has \textit{16} choices in \textit{SMS}, while it has $16 \times framesize$ different choices in our approach to mitigate the interference, which explains the large difference in \textit{McP} between our approach and \textit{SMS}.
\subsubsection{Power Consumption}
The mean power consumption denoted by \textit{mPC} of each \textit{WBAN} versus the number of coexisting \textit{WBAN}s ($\Omega$) for \textit{DAIL}, \textit{CHIM} and \textit{SMS} are compared in \textbf{Figure \ref{energydcs}}. As seen in the figure, both \textit{DAIL} and \textit{CHIM} consistently have lower \textit{mPC} than \textit{SMS} for all $\Omega$ values. Such performance advantage is attributed to reduction in the number of medium access collisions which lead to a fewer number of retransmissions and hence high energy savings. Meanwhile, in \textit{SMS}, \textit{mPC} is high due to the medium access collisions resulting from the large number of sensors that compete for the available channels (\textit{16 channels}) and hence the power consumption grows accordingly.
\section{Conclusions}
In this paper, we have presented a distributed approach to minimize the impact of inter-\textit{WBAN} interference through dynamic channel hopping based on Latin rectangles. Furthermore, the proposed approach opts to reduce the overhead resulting from channel hopping and lowers the transmission delay as well as saves the power resource at both \textit{sensor}- and \textit{WBAN}-levels. Specifically, we propose two schemes for channel allocation and medium access scheduling to diminish the probability of inter-\textit{WBAN} interference. The first scheme, namely, \textit{DAIL}, suits high density of coexisting \textit{WBAN}s and involves overhead due to frequent channel hopping at the coordinator and the sensors. \textit{DAIL} combines the channel and time-slot hopping to lower the probability of collisions among transmission of sensors in the high density of coexisting \textit{WBAN}s. Each distinct \text{WBAN}'s coordinator autonomously picks an orthogonal Latin rectangle and assigns its individual sensors unique transmission patterns. The second scheme, namely, \textit{CHIM}, takes advantage of the relatively lower density of coexisting \textit{WBAN}s to save power by hopping among channels only when interference is detected at the level of the individual nodes. We have further presented an analytical model that derives the collision probability, network throughput, and transmission delay. Simulation results show that \textit{DAIL} and \textit{CHIM} outperform other competing schemes.
\begin{wrapfigure}{l}{25mm} 
    \includegraphics[width=1in,height=1.25in,clip,keepaspectratio]{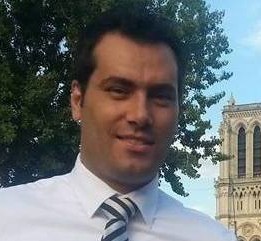}
  \end{wrapfigure} \par
\textbf{Mohamad Jaafar Ali} Mr. Ali is currently a Ph.D. student in the department of computer science at university of Paris Descartes (Paris 5). He received his M.Sc. degree in computer science from university of Grenoble (Grenoble 1), France. His research interests include architectures and communication protocols for mobile wireless and wireless body area networks (WBANs) and Internet of Things (IoT).\par
\begin{wrapfigure}{l}{25mm} 
    \includegraphics[width=1in,height=1.25in,clip,keepaspectratio]{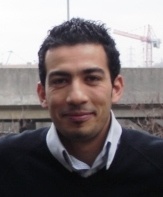}
\end{wrapfigure}\par
\textbf{Hassine Moungla} Mr. Moungla is currently working as CNRS Researcher at Institut Mines Telecom - Telecom Sud Paris located at CEA Saclay Nano-Innov pole. He is an associate professor at the university Paris Descartes (Paris 5) and a member of (Paris 5) Laboratory (LIPADE) since october 2008. He received his Ph.D. degree in computer science from university of Paris 13 on 2006. His research interests include Wireless Sensor Networking for medical applications, Wireless Sensor Networking, Multimedia over Wireless, QoS in WLAN/WSN, Middelware for 4G Mobile and Sensor Networks. \par
\begin{wrapfigure}{l}{25mm} 
    \includegraphics[width=1in,height=1.25in,clip,keepaspectratio]{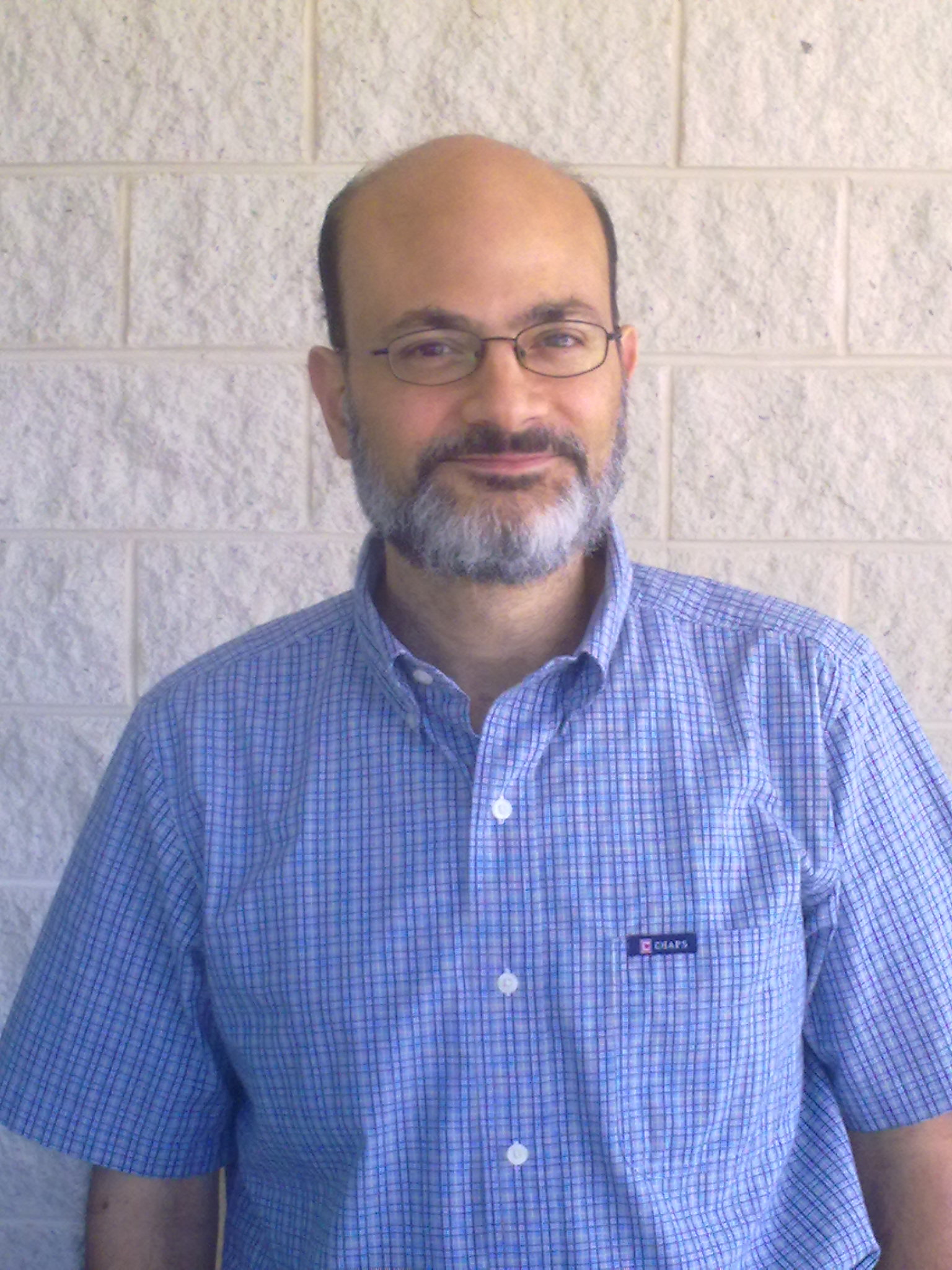}
\end{wrapfigure}\par
\textbf{Mohamed Younis} Mr. Younis is currently an associate professor in the department of computer science and electrical engineering at the university of Maryland Baltimore County (UMBC). He received his Ph.D. degree in computer science from New Jersey Institute of Technology, USA. Before joining UMBC, he was with the Advanced Systems Technology Group, an Aerospace Electronic Systems R$\&$D organization of Honeywell International Inc. While at Honeywell he led multiple projects for building integrated fault tolerant avionics and dependable computing infrastructure. He also participated in the development of the Redundancy Management System, which is a key component of the Vehicle and Mission Computer for NASA's X-33 space launch vehicle. Dr. Younis' technical interest includes network architectures and protocols, wireless sensor networks, embedded systems, fault tolerant computing, secure communication and distributed real-time systems. He has published over 240 technical papers in refereed conferences and journals. Dr. Younis has six granted and two pending patents. In addition, he serves/served on the editorial board of multiple journals and the organizing and technical program committees of numerous conferences. Dr. Younis is a senior member of the IEEE and the IEEE communications society. \par
\begin{wrapfigure}{l}{25mm} 
    \includegraphics[width=1in,height=1.25in,clip,keepaspectratio]{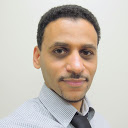}
\end{wrapfigure}\par
\textbf{Ahmed Mehaoua} Mr. Mehaoua received the M.Sc. and Ph.D. degrees in computer science from the University of Paris, Paris, France, in 1993 and 1997, respectively. He is currently a Full Professor of computer networking in the Faculty of Mathematics and Computer Science, at University of Paris Descartes, Paris, France. He is also the Head of the Department of Multimedia Networking and Security at the LIPADE, a governmental computer science research center in Paris, France. His research interests include resource optimization, security and anomaly detection in wired and wireless networks.\par
%\begin{thebibliography}{00}
 
%\end{thebibliography}
\end{document}